\documentclass[twocolumn,amsfonts,showpacs,superscriptaddress,nofootinbib]{revtex4-1}
\usepackage{amsfonts,amssymb,amsmath}
\usepackage{lipsum, babel}
\usepackage{mathtools}
\usepackage[utf8]{inputenc}
\usepackage[pdftex,breaklinks=true,colorlinks=true]{hyperref}
\usepackage[pdftex]{graphicx}
\usepackage{epstopdf}
\usepackage{leftidx}
\graphicspath{{Pictures/}}

\newcommand{\bc}{\begin{center}}
\newcommand{\ec}{\end{center}}
\def\ba#1{\begin{array}{#1}\displaystyle}
\newcommand{\ea}{\end{array}}

\newcommand{\beq}{\begin{equation}}
\newcommand{\eeq}{\end{equation}}
\newcommand{\beqa}{\begin{eqnarray}}
\newcommand{\eeqa}{\end{eqnarray}}

\newcommand{\n}{\nonumber\\}
\newcommand{\bi}{\begin{itemize}}
\newcommand{\ei}{\end{itemize}}

\newcommand{\p}{\partial}

\newcommand{\dd}{{\rm d}}

\newcommand{\dr}{\mathrm{dr}}

\newcommand{\ttbar}{$T\bar{T}$}

\newcommand{\ii}{{\rm i}}

\newcommand{\tot}{{\rm tot}}

\newcommand{\eff}{{\rm eff}}

\def\eqref#1{(\ref{#1})}

\newcommand{\cL}{{\cal L}}
\newcommand{\bcL}{\bar{\cal L}}

\usepackage{amsmath}	
\begin{document}


\title{Thermal transport in $T\bar{T}$-deformed conformal field theories: from integrability to holography}

\author{Marko Medenjak}
\affiliation
{Institut de Physique Th\'eorique Philippe Meyer, \'Ecole Normale Sup\'erieure, \\ PSL University, Sorbonne Universit\'es, CNRS, 75005 Paris, France}

\author{Giuseppe Policastro}
\affiliation{Laboratoire de Physique de l’Ecole Normale Sup\'erieure, CNRS, Universit\'e PSL, Sorbonne 
Universit\'es, Universit\'e Pierre et Marie Curie, 24 rue Lhomond, 75005 Paris, France}

\author{Takato Yoshimura}
\affiliation{Department of Physics, Tokyo Institute of Technology, Ookayama 2-12-1, Tokyo 152-8551,
Japan}
\affiliation
{Institut de Physique Th\'eorique Philippe Meyer, \'Ecole Normale Sup\'erieure, \\ PSL University, Sorbonne Universit\'es, CNRS, 75005 Paris, France}

\begin{abstract}
In this paper we consider the energy and momentum transport in (1+1)-dimension conformal field theories (CFTs) that are deformed by an irrelevant operator $T\bar{T}$, using the integrability based generalized hydrodynamics, and holography. The two complementary methods allow us to study the energy and momentum transport after the in-homogeneous quench, derive the exact non-equilibrium steady states (NESS) and calculate the Drude weights . Our analysis reveals that all of these quantities satisfy universal formulae regardless of the underlying CFT, thereby generalizing the universal formulae for these quantities in pure CFTs. We also compute the exact momentum diffusion constant using the integrability-based method, and confirm that it agrees with the conformal perturbation. These fundamental physical insights have important consequences for our understanding of the $T\bar{T}$-deformed CFTs. First of all, they provide the first check of the $T\bar{T}$-deformed $\mathrm{AdS}_3$/$\mathrm{CFT}_2$ correspondence from the dynamical standpoint. And secondly, we are able to identify a remarkable connection between the $T\bar{T}$-deformed CFTs and reversible cellular automata.
\end{abstract}

\maketitle
\section{Introduction}
Spreading of energy is a central topic of thermodynamics, which has been scrutinized already in the Fourier's work on the energy flow in materials, and, and Stefan's and Boltzmann's work on the black body radiation. Stefan-Boltzmann law states that the energy current between two regions at different temperatures is simply proportional to the difference of the temperatures to the power $d$, where $d$ is the dimension of the system. Rather surprisingly, it was recently noticed that Stefan-Boltzmann law exactly describes the energy transport in conformal field theories (CFTs) \cite{cardy2010ubiquitous,bernard2012energy}.

In this paper we shall take a step forward, by studying how a departure from the strict CFT limit affects the spreading of energy and momentum. In particular we will be concerned with the effects of irrelevant interaction on the dynamics of (1+1)-dimensional CFTs. Such situation is especially pertinent in gapless systems where the asymptotic approach to the low-energy sector is controlled by irrelevant operators. Indeed, a systematic consideration of their effects culminated in the birth of non-linear Luttinger liquid theory, in which the nonlinear dispersion relation gives rise to important ramifications of conventional Luttinger liquid theory \cite{Imambekov228,1367-2630-17-10-103003}.  While in general irrelevant perturbations are hard to handle, it has recently been discovered that a special classes of deformed theories, so called $T\bar{T}$-deformations \cite{Zamolodchikov:2004ce,Dubovsky2012,Caselle2013,Cavagli2016,SMIRNOV2017363,Kraus2018,Cardy2018,Conti2019} (for a review, see \cite{jiang2019lectures}), are exactly solvable. $T\bar{T}$-deformations have a number of remarkable properties, and using their structure will allow us to address the questions associated with transport and thermodynamics from two complementary perspectives: integrability and holography.

Recently, we have witnessed rapid developments in our understanding of the dynamical phenomena in interacting integrable systems, which have been mainly driven by the advent of generalized hydrodynamics (GHD) \cite{PhysRevX.6.041065,PhysRevLett.117.207201} (see \cite{10.21468/SciPostPhysLectNotes.18} for a review). Integrable systems are characterized by possessing an infinite number of conserved charges in the infinite size limit, which presents an apparent obstacle in using the usual hydrodynamic description. The pursuit of hydrodynamics of integrable systems resulted in the discovery of GHD. Due to its integrable foundations, GHD allows for analytical treatment, which goes well beyond the conventional hydrodynamics  \cite{10.21468/SciPostPhys.6.2.023,PhysRevX.10.011054,PhysRevE.101.060103,PhysRevLett.125.070602,10.21468/SciPostPhys.9.3.040}. In particular, it led to numerous exact results on transport coefficients \cite{SciPostPhys.3.6.039,PhysRevLett.121.160603,De_Nardis_2019}, operator spreading \cite{Gopalakrishnan2018}, entanglement spreading \cite{Alba7947}, integrability breaking \cite{PhysRevB.101.180302,10.21468/SciPostPhys.9.4.044,lopezpiqueres2020hydrodynamics,bastianello2020thermalisation}, etc. Furthermore, predictions of GHD have provided appropriate description of experimental observations which go beyond the standard hydrodynamics \cite{PhysRevLett.122.090601,malvania2020generalized}. In this paper we shall fully exploit the machinery of GHD to explore the out-of-equilibrium dynamics of $T\bar{T}$-deformed CFTs.

On the other side, we have the holographic correspondence, which maps a class of strongly-coupled CFTs to a weakly coupled dual gravitational theory on the  asymptotically anti-de Sitter space background (a space of negative constant curvature); the dual theory can be treated semiclassically in the limit of large central charge of the CFT \cite{Maldacena:1997re,Witten:1998qj,Gubser:1998bc}. The correspondence has proven very powerful especially in the hydrodynamic regime, where it led to many new insights concerning the transport properties of strongly coupled systems at a critical point, {\it e.g.} universal properties of certain transport coefficients \cite{Policastro:2001yc,Kovtun:2004de}, effects of quantum anomalies on transport \cite{Son:2009tf}, effective theory for dissipative fluids \cite{Haehl:2018lcu,Glorioso:2018wxw} (for a review see \cite{Ammon:2015wua,CasalderreySolana:2011us}). 
Since in (1+1)-dimensional CFTs the hydrodynamic behavior is completely determined by the symmetry, the holographic correspondence cannot provide any new insights in this regard. The behavior of the $T\bar{T}$-deformed CFTs is, however, markedly different, opening an opportunity to study physical phenomena associated with the breaking of conformal invariance and to establish the first connection between holography and GHD.

In this manuscript we provide the exact solutions of the partitioning protocol, where two $T\bar{T}$-deformed CFTs at different temperatures and different momenta are joined together and let to evolve unitarily. The partitioning protocol is one of the simplest setups for studying the  dynamics, and has been analyzed in many models, ranging from higher dimensional CFTs to hard rod models \cite{Bhaseen:2015aa,Doyon_2017}.  It was also realized recently that the protocol can be used to compute the Drude weights, i.e. the quantities that measure the strength of ballistic transport, using a linear response argument \cite{SciPostPhys.3.6.039,IN_Drude}. Intriguingly, we find that in $T\bar{T}$-deformed CFTs both the energy and momentum NESS currents are universal, which also implies the universality of the energy and momentum Drude weights. The agreement of these quantities computed by integrability and holography provides, as far as we are aware, the first instance where the $T\bar{T}$-deformed AdS/CFT correspondence in (1+1)-dimensions is validated from a dynamical point of view. Further, we show that $T\bar{T}$-deformation gives rise to diffusive corrections that are absent in a pure CFT. We argue, however, that they are suppressed in the large-$c$ holographic $T\bar{T}$-deformed CFT.  On the hydrodynamic level we also observe a curious connection between the $T\bar{T}$-deformation and the reversible cellular automaton 54 (RCA 54) \cite{Bobenko1993}. It turns out that RCA 54 can be understood as a discrete space-time version of a $T\bar{T}$-deformed CFT with hard-core interaction between the same species of particles.

The paper is structured as follows. In the first two sections we discuss the basics of energy transport in one dimension and $T\bar{T}$-deformation. This is followed by overviews of main results before discussing the details of them. We then start the main part with introducing the integrability-based treatment of  $T\bar{T}$-deformation in the scope of thermodynamic Bethe ansatz (TBA) and GHD. Using these results we shall obtain the Drude weights, diffusive corrections and the nonequilibrium steady states (NESS) following the partitioning protocol, and discuss the connection with the RCA 54. To provide better understanding of the structure of \ttbar-deformed CFTs, we then analyze two concrete examples; \ttbar-deformation of the critical Ising CFT and \ttbar-deformed Liouville CFT, which is an interacting CFT. Having obtained the main results by making use of GHD, we then move on to the holographic parts. First we briefly review the holographic approach to transport, after which we derive the Drude weights, solve for NESS in $T\bar{T}$-deformed holographic CFTs. Finally as a consistency check, we compute the momentum diffusion constant using conformal perturbation up to the second order in the deformation parameter before closing the paper with an outline of future directions. This paper is a longer and more detailed version
of the companion paper \cite{prl}.
\vspace{0.2cm}

\section{Energy transport}
Stefan Boltzmann law $j=\sigma_\mathrm{SB} T^{d+1}$, describing the black body radiation, relates the energy flux $j$ in $d$-dimensions to its temperature $T$, where $\sigma_\mathrm{SB}$ is the Stefan Boltzmann constant. The thermodynamic derivation of the law by Boltzmann, however, relies on the tracelessness of the stress-energy tensor $T^\mu_\mu=0$.

In recent years a more general setup called the partitioning protocol has been extensively studied in (1+1) dimensional CFTs. In this setup the left and the right sides of the infinitely large system are initialized at different temperatures, $T_L$ and $T_R$. In the long time limit the system relaxes to the ever expanding NESS. In the scope of the partitioning protocol the exact NESS energy and charge currents, as well as their fluctuations were obtained for CFTs \cite{bernard2016conformal}. The full space-time NESS average profile $\langle j_E(x,t)\rangle$ of the energy current operator $j_E$ is particularly simple; it is given by $\langle j_E(x,t)\rangle=\mathtt{j}_\mathrm{NESS}\vartheta(x-|t|)$, where $\mathtt{j}_\mathrm{NESS}=\pi c(T^2_L-T^2_R)/12$ with the central charge $c$, and $\vartheta(x)$ the step function. Hence the NESS energy current takes a constant value $\mathtt{j}_\mathrm{NESS}$ inside of the light cone $|t|<x$ and is zero otherwise. In particular, the NESS at $x=0$ can be thought of as a boosted thermal state $\rho\sim e^{-\beta H+\nu P}$ with the rest frame inverse temperature $\beta=\sqrt{\beta_R\beta_L}$ and the boost parameter $\tanh\nu=(\beta_L-\beta_R)/(\beta_L+\beta_R)$. Such a simple profile is a consequence of chiral separation in pure CFTs; elementary excitations in CFTs are simply the right and the left moving chiral modes, which do not interact. Consequently right (resp. left) movers contributing to the NESS are thermalized with respect to the left (resp. right) bath. As expected, if one of the temperatures, say $T_R$, is set to $0$, we reproduce the Stefan Boltzmann law exactly \cite{cardy2010ubiquitous}.

The partitioning protocol can also be used to obtain linear response transport coefficients such as the Drude $D_{ij}$ and diffusion $\mathfrak{D}_{i}^{\,\,j}$ coefficients. Drude weights characterize the persistence of the current in the system
\begin{equation}
    D_{ij}=\lim_{t\to\infty}\int_{-t}^t\frac{\dd s}{2t}\int_\mathbb{R}\dd x\langle j_i(x,s)j_j(0,0)\rangle^c,
\end{equation}
where $j_i(x,t)$ is the current density associated with the conserved density $q_i$, while the connected correlation function $\langle o_1(x,t)o_2(0,0)\rangle^c$ is evaluated in the stationary state at some temperature. From the point of view of partitioning protocol, $D_{ij}$ can be interpreted as the increase of the current $j_i$, generated due to the infinitesimal bias of the chemical potential $\delta\mu_j$ pertaining to the conserved quantity $q_j$ \cite{IN_Drude}
\begin{equation}
    D_{ij}=\lim_{t\to\infty}\frac{1}{2t}\frac{\int_{\mathbb{R}}\dd x\left<j_i(x,t)\right>_{\delta \mu_j}}{\delta \mu_j}.
\end{equation}
Diffusion constants, on the other hand, control the broadening of the ballistic trajectories. It is useful to consider the Onsager matrix $\mathfrak{L}=\mathfrak{D}C$ instead of the bare diffusion constants $\mathfrak{D}_i^{\,\,j}$, which reads \cite{De_Nardis_2019}
\begin{equation} \label{Onsager}
    \mathfrak{L}_{ij}=\int_\mathbb{R}\dd t\left(\int_\mathbb{R}\dd x\langle j_i(x,t)j_j(0,0)\rangle^c-D_{ij}\right),
\end{equation}
where $C_{ij}=\int_\mathbb{R}\dd x\langle q_i(x,0)q_j(0,0)\rangle^c$ is the susceptibility matrix. 
The Onsager matrix has the physical meaning of a matrix of conductivities; the frequency-dependent conductivity (at zero momentum) has generically the form \cite{Bernard_2016} $\sigma_{ij}(\omega) = D_{ij} \delta(\omega) +\tilde \sigma_{ij}(\omega)$, where $\tilde \sigma_{ij}$ is a smooth function at $\omega=0$ and $\tilde \sigma_{ij}(\omega=0) = \mathfrak{L}_{ij}$ . The finite part of the conductivity can be recovered also from the well-known Green-Kubo formula \cite{Kubo:1957mj,Green:1952}: 

\begin{equation}\label{kubo}
{\mathfrak L}_{ij} = - \lim_{\omega\to 0} \lim_{k\to 0} \frac{1}{\omega} \textrm{Im} \, {\mathcal G}^R_{ij} (\omega,k)  \,,
\end{equation}where $\mathcal{G}^R_{ij}(\omega,k) = - i \int \dd t \dd x\, e^{ikx-i\omega t} \theta(t) \langle [j_i(x,t),j_j(0,0)] \rangle$ is the retarded current-current correlator. 

\section{$T\bar{T}$-deformation}
$T\bar{T}$-deformation of the (1+1)-dimensional quantum field theory with Lagrangian $\mathcal{L}^{(0)}$ is obtained through the sequence of infinitesimal changes $\mathcal{L}^{(\sigma)}\mapsto\mathcal{L}^{(\sigma+\delta\sigma)}$ as
\begin{equation}
\mathcal{L}^{(\sigma+\delta\sigma)}=\mathcal{L}^{(\sigma)}+\frac{\delta\sigma}{2}\det  T_{\mu\nu}.
\end{equation}
The trajectory is induced by a composite operator $\det T_{\mu\nu}=-\frac{1}{\pi^2}(T\bar{T}-\Theta^2)$, where $T(z,\bar{z})=-2\pi T_{zz}$, $\bar{T}(z,\bar{z})=-2\pi T_{\bar{z}\bar{z}}$, and $\Theta(z,\bar{z})=-2\pi T_{z\bar{z}}$ are the components of the stress-energy tensor $T_{\mu\nu}$. 
Importantly, $T_{\mu\nu}$ depends on the value $\sigma$. A salient feature of this deformation is that independently of the original theory $\mathcal{L}^{(0)}$, a finite-volume spectrum $E_n(R,\sigma)$ can be obtained from the undeformed energies $E_n(R,0)$, by solving the Burgers equation \cite{Zamolodchikov:2004ce}
\begin{equation}\label{burgers}
    \p_\sigma E_n(R,\sigma)=E_n(R,\sigma)\p_RE_n(R,\sigma)+\frac{1}{R}P^2_n(R),
\end{equation}
where $R$ is the volume of the (compactified) system. Note that the momenta $P_n(R)=2\pi p_n/R,\,p_n\in\mathbb{Z}$ remain undeformed. This suggests that if we know the spectrum of the undeformed theory, which is the case in CFTs and integrable systems, we can immediately obtain that of the deformed system. For instance, solving the Burgers equation, in the infinite volume, the free energy of a \ttbar-deformed CFT with the central charge $c$ at temperature $\beta=R$, which in fact equals $f=E_0(\beta,\sigma)/\beta$, can be computed exactly, reading
\begin{equation} \label{free-energy}
   f=-\frac{1}{\sigma}\left(1-\sqrt{1-\frac{\pi\sigma c}{3\beta^2}}\right).
\end{equation}
Observe that when $\sigma>0$, the free energy becomes complex when $T$ exceeds the Hagedorn temperature $T_\mathrm{H}=\sqrt{3/\pi\sigma c}$, signifying the peculiar UV behavior of \ttbar-deformed CFTs. It is also important to note that the \ttbar-deformation keeps the structure of conserved charges intact: for instance in \ttbar-deformed CFTs, conserved charges are given by the deformed KdV charges, which are still local \cite{10.21468/SciPostPhys.7.4.043}.

There is another consequence of the $T\bar{T}$-deformation, which is central in the study of thermodynamics and hydrodynamics. Consider a scattering involving asymptotic incoming particles with a set of four-momenta $\{p^\mu_a\}$ and outgoing particles with $\{q^\mu_a\}$, where $p^\mu=(E,p)$. Then the S-matrix of the deformation is given by \cite{Dubovsky2017,Cardy2019}
\begin{align}
    S^{(\sigma)}(\{p^\mu_a\},\{q^\mu_a\})&=e^{\ii \sigma\sum_{a<b}\epsilon_{\mu\nu}p^\mu_ap^\nu_b+\ii \sigma\sum_{a<b}\epsilon_{\mu\nu}q^\mu_aq^\nu_b} \n
   &\quad\times S^{(0)}(\{p^\mu_a\},\{q^\mu_a\}),
\end{align}
where $S^{(0)}(\{p^\mu_a\},\{q^\mu_a\})$ is the S-matrix of the original theory. The deformation at the level of S-matrix can be inferred from a geometric interpretation of the \ttbar-deformation, or similarly, by thinking of it as a field-dependent coordinate transformation \cite{Cardy2019}. On the other hand, S-matrix can be  obtained by resorting to the identification of the \ttbar-deformation and the coupling of the undeformed theory to the Jackiw-Teitelboim gravity \cite{Dubovsky2017,Dubovsky2018}, hence the name gravitational dressing. This modification is rather simple for integrable systems with a two-body S-matrix $S^{(\sigma)}(\theta)$. In the deformed theory, S-matrix is obtained by the multiplication with the CDD factor $\Sigma(\theta)$: $S^{(\sigma)}(\theta)=e^{\ii \Sigma(\theta)}S^{(0)}(\theta)$. 
The CDD factor is $\Sigma(\theta)=\sigma m^2\sinh\theta$ when the undeformed theory is massive, while $\Sigma(\theta)=\frac{\sigma}{2}M^2e^\theta=-\sigma p_+(\theta_1)p_-(\theta_2)$ with $\theta=\theta_1-\theta_2$ and $p_\pm(\theta)=\pm\frac{M}{2}e^{\pm\theta}$ when perturbing a massless theory. Here $M$ is the energy scale that controls the crossover from ultraviolet to infrared. We can formally apply the same logic to CFTs, even though the notion of S-matrix is not well-defined in these cases. This is achieved by regarding CFTs as massless integrable models following the works by Bazhanov, Lukyanov, and Zamolodchikov \cite{Bazhanov1996}. Later we will implement this idea by focusing on CFTs that can be described by two non-linear integral equations.\\

\begin{figure}[h!]
\centering
\includegraphics[width=7cm]{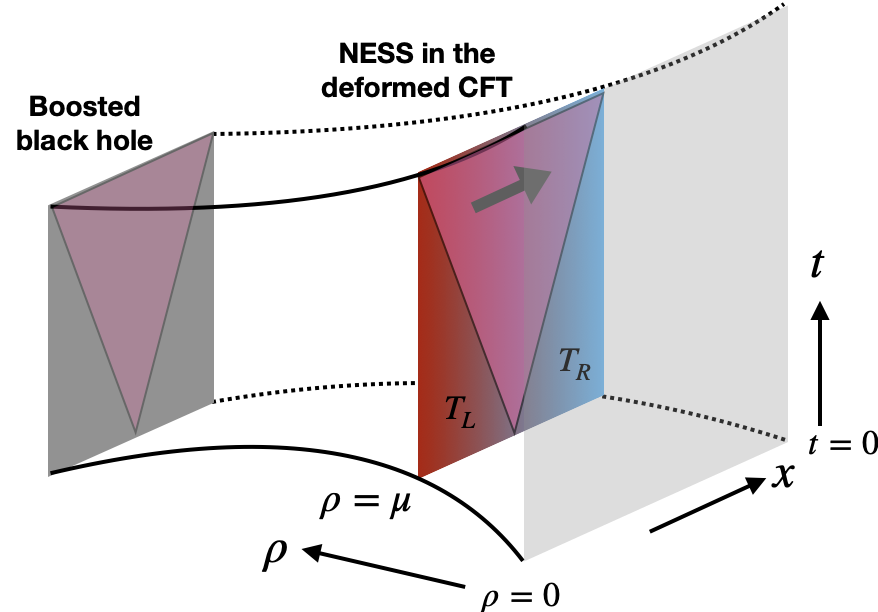}
\caption{Correspondence between the AdS with the finite radius and the $T\bar{T}$-deformed CFT.}
\label{ads}
\end{figure}

While the solvability of the $T\bar{T}$-deformation is not manifest in the holographic setting, it might be reflected by the simplicity of the deformed correspondence. In \cite{McGough:2016lol} it was conjectured that the deformation corresponds to imposing Dirichlet boundary conditions on the gravity fluctuations on a surface at a finite value of the radial coordinate $\rho$. The evidence for the conjecture is that the equation of state derived holographically with this prescription is in agreement with the field theory result; however \cite{Guica:2019nzm} argued for a different approach. By analysing the phase space of solutions, they found that the correct prescription is to consider a modified mixed boundary condition for the graviton at the boundary. They also showed that this prescription is equivalent to the finite-cutoff one in the case where other fields apart from the graviton are not sourced (but they can have an expectation value). This is the case we consider since we take a setup where gravity is the only field, so we will still refer to 
the finite-cutoff prescription. It is important to mention though  that the interpretation as mixed boundary condition allows for both signs of the deformation, whereas the finite cutoff can only make sense for one sign, as we will see in section \ref{sec:holographic}. Fig. \ref{ads} illustrates the $AdS$ geometry that we have in mind.  

\section{Main results}
Before proceeding any further, let us present the main results. We consider an arbitrary \ttbar-deformed CFT with the central charge $c$ and the deformation parameter $\sigma$ at a finite temperature $T=1/\beta$. Firstly let us focus on the energy and momentum NESS currents, which emerge after joining two thermal baths of the systems with different temperatures $T_L$ and $T_R$. From two independent computations, one based on integrability and another that makes use of holography, they turn out to be given by
\begin{align} \label{NESScurrent_ttbar}
    \langle j_E\rangle_\mathrm{NESS}&=\frac{\pi c }{12}e_{RL}\left(\tilde{T}^2_L-\tilde{T}^2_R\right),\n
    \langle j_P\rangle_\mathrm{NESS}&=\frac{\pi c }{12}e_{RL}\left(\tilde{T}^2_L+\tilde{T}^2_R-\frac{\pi c\sigma}{6}\tilde{T}_L^2\tilde{T}_R^2\right),
\end{align}
where
\begin{equation}
    \tilde{T}(T)=\frac{2T}{1+\sqrt{1-\frac{\pi\sigma cT^2}{3}}},\quad e_{RL}=\frac{1}{1-\left(\frac{\pi\sigma c}{12}\right)^2\tilde{T}_L^2\tilde{T}_R^2}
\end{equation}
with $\tilde{T}_{R,L}=\tilde{T}(T_{R,L})$. It is remarkable that such general formulae hold even away from critically. We emphasize that {\it a priori} there is no reason why the NESS currents computed by these two methods have to agree, because in general the dynamics of integrable CFTs (e.g. minimal models) and holographic CFTs can qualitatively differ \cite{Asplund2015}. However our analysis reveals that, in fact, as far as the energy and momentum transport are concerned, physics seems to be rather universal. As we discuss below, there is strong evidence that such universality also persists in the momentum diffusion constant. We suspect that this phenomenon is strongly tied to the fact that, at least in the classical case, the energy, momentum, and pressure evolve without being mixed with other higher-spin conserved charges \cite{jorjadze2020canonical}. Since we are dealing with the hydrodynamic behavior of the system, which is presumably classical, it is reasonable that such a decoupling also occurs at the level of mean values, which amounts to the  universal  NESS formulae. We also note that these results reproduce the known universal formulae in pure CFTs \cite{bernard2012energy}
\begin{align} \label{NESScurrent_CFT}
    \langle j_E\rangle_\mathrm{NESS}&=\frac{\pi c }{12}\left(T^2_L-T^2_R\right),\n
    \langle j_P\rangle_\mathrm{NESS}&=\frac{\pi c }{12}\left(T^2_L+T^2_R\right).
\end{align}
It should be stressed that our derivation based on integrability does not appeal to any technique of CFT, hence serves as a yet another derivation of these CFT formulae.
The universality of NESS currents automatically implies that of Drude weights as well. Let us recall that in a pure CFT they are obtained by \eqref{NESScurrent_CFT}, and read \cite{bernard2012energy}
\begin{equation}\label{drudeCFT}
    D_{EE}=\frac{\pi c}{3v}T^3,\quad D_{PP}=\frac{\pi c v}{3}T^3
\end{equation}
where $v$ is the Fermi velocity or the sound velocity. It is related to the dispersion relation at the quantum critical point $E\sim v|p|$ and determined by the details of the underlying microscopic model (Luttinger liquid, quantum Hall edge states etc). In this article we set it to $v=1$. Our computations reveal that also for \ttbar-deformed theories Drude weights take exceedingly simple universal forms in terms of the energy density $\mathtt{e}$ and pressure  $\mathtt{p}$
\begin{equation}
       D_{EE}= \frac{\mathtt{e}+\mathtt{p}}{\beta}=\frac{\pi c}{3 v_c}T^3,\, D_{PP}=\left( \frac{\mathtt{p}}{\mathtt{e}} \right)^2 D_{EE}=\frac{\pi cv_c}{3}T^3,
\end{equation}
where $v_c$ is the generalized sound velocity depending only on the deformation parameter $\sigma$ and the central charge $c$
\begin{equation}\label{effective-v}
   v_c=\sqrt{1-\frac{\pi\sigma cT^2}{3}}.
\end{equation}
This is a rather natural generalization of the Drude weight formulae in pure CFTs; the only change induced by the \ttbar-deformation is the modification of the sound velocity, which is reminiscent of how the spectrum of CFTs is altered by the deformation. This also confirms previous perturbative results \cite{Cardy_2016,Bernard_2016,McGough:2016lol}.

One can show that in 2d CFTs diffusion is absent in general. Importantly, we will prove that the deformation gives rise to the finite momentum diffusion constant, which, remarkably, in generic \ttbar-deformed CFTs, read
\begin{equation} \label{LPP}
  \mathfrak{L}_{PP}=\frac{\sigma^2}{2}v_c D_{EE}^2=\frac{\sigma^2}{2}\frac{D_{PP}^2}{v^3_c}
\end{equation}
It turns out that the momentum diffusion, which is also called the bulk viscosity, is related to the entropy density $s$ in a simple fashion
\begin{equation}
    \mathfrak{L}_{PP}=\frac{\pi c}{6\beta^5}\sigma^2 s, \quad s=\frac{\pi c}{3v_c\beta}.
\end{equation}
We also support this computation by carrying out the conformal perturbation up to the second order in the deformation parameter. Note that the energy diffusion is still absent in \ttbar-deformed CFTs due to their Lorentz invariance \cite{De_Nardis_2019}. However, we will argue that in the classical gravity also the momentum diffusion vanishes, implying that it originates from the quantum gravity corrections that are suppressed by powers of $1/c$.


Lastly we also point out a connection between \ttbar-deformed CFTs and an integrable cellular automaton model called the reversible cellular automaton 54 (RCA 54). To be more precise, the energy density of \ttbar-deformed CFTs and the soliton density of the RCA 54, both of which are denoted by $\rho_\pm$ here, satisfy the same hydrodynamic equations
\begin{equation}\label{hydrobobenko}
\p_t\rho_\pm+\p_x(v^\eff_\pm\rho_\pm)=0,\quad v^\eff_\pm=\frac{\pm1+\sigma(\rho_+-\rho_-)}{1+\sigma(\rho_++\rho_-)},
\end{equation}
whereby the identification of the energy quanta in \ttbar-defomed CFTs and solitons in the RCA 54 is established. 
\section{Thermodynamics of \ttbar-deformed CFTs}
Here we shall elaborate on the universal structure of the \ttbar-deformed CFTs from the integrability point of view. The discussion will rest upon the assumption that the model can be formulated {\it \`a la} Bazhanov-Lukyanov-Zamolodchikov with the phase shift $T$, which holds for numerous CFTs, such as minimal models ($c<1$) and the Liouville CFT ($c\geq25$). Later we illustrate how the general idea works for two of these models: critical Ising model and the Liouville CFT. Below, for simplicity, we also focus on the case where the CFT can be described by two non linear integral equations (NLIE) only, one for the left movers and one for the right movers. The NLIEs for CFTs in a generalized Gibbs ensemble (GGE) $\rho\sim e^{-\beta H+\nu P-W}$, where $W=\sum_{i=2}^\infty\beta^iQ_i$ with higher conserved charges $Q_i$ and associated Lagrange multipliers $\beta^i$, then read
\begin{equation}\label{nlieGGE1}
    \varepsilon_\pm(\theta)=(\beta\mp\nu) E_\pm(\theta)+w_\pm(\theta)-T\star L_\pm(\theta),
\end{equation}
where $E_\pm(\theta)=Me^{\pm\theta}/2$, $L_\pm(\theta)=\log(1+e^{-\varepsilon_\pm(\theta)})$, and $w_\pm(\theta)=\sum_{i=2}^\infty\beta^ih_{i,\pm}(\theta)$, with $h_{i,\pm}(\theta)\propto e^{\pm i\theta}$ one-particle eigenvalues of $Q_i$. Here $\star$ stands for the convolution: $T*L(\theta)=\int\dd\theta'T(\theta,\theta')L(\theta')$, and the phase shift $T(\theta)$ depends on the parameter. For their asymptotics we assume the usual ones: $\varepsilon_\pm(\theta)\to(\beta\mp\nu)E_\pm(\theta)+w_\pm(\theta)$, when $\theta\to\pm\infty$ while $\varepsilon_\pm(\theta)\to0$ for $\theta\to\mp\infty$. Note that the two equations are decoupled without the \ttbar-deformation, hence we can treat them separately. Also $\varepsilon_+(\theta)=\varepsilon_-(-\theta)$ in a thermal state $\nu=w_\pm(\theta)=0$. In the absence of $R-L$ scattering and higher charges, the scaling of $M\mapsto sM$ merely shifts the rapidity $\theta\mapsto\theta+\log s$. This implies a lack of characteristic scale of the system, which is a hallmark of conformal invariance. The central charge of the system is given by the scaling function $\tilde{c}(\beta,\nu)=\tilde{c}_+(\beta,\nu)+\tilde{c}_-(\beta,\nu)$ where
\begin{equation}\label{centralcharge}
     \tilde{c}_\pm(\beta,\nu)=\frac{3(\beta\mp\nu) M}{2\pi^2}\int_{-\infty}^\infty\dd \theta e^{\pm\theta} L_\pm(\theta),
\end{equation}
for $\nu=w_\pm(\theta)=0$ (i.e. thermal state).
Invoking the standard dilogarithm calculus in TBA, one can usually carry out the integration, recovering the central charge of the CFT $\tilde{c}_+(\beta,0)=c$ in terms of the model parameters that also enter into $T$. Note that in the presence of higher charges the scaling function no longer equals $c$.

Now, let us turn on the \ttbar-deformation, which induces the $R-L$ scatterings, resulting in the phase shift $\tilde{T}_{\pm\mp}(\theta,\theta')=\frac{\sigma M^2}{8\pi}e^{\pm(\theta-\theta')}=\tilde{T}^\mathrm{T}_{\pm\mp}(\theta,\theta')$. The TBA equations for the \ttbar-deformed CFT then reads
\begin{equation}
    \varepsilon_\pm(\theta)=(\beta\mp\nu) E_\pm(\theta)+w_\pm(\theta)-T\star L_\pm(\theta)-\tilde{T}_{\pm\mp}\star L_\mp(\theta),
\end{equation}
which can be conveniently rewritten as
\begin{equation}\label{nlieGGE2}
    \varepsilon_\pm(\theta)=\left(\beta\mp\nu-\frac{\pi \sigma}{6(\beta\pm\nu)}\tilde{c}_{\mp}\right)E_\pm(\theta)+w_\pm(\theta)-T\star L_\pm(\theta).
\end{equation}
\ttbar-deformation only affects the UV property of the CFT; the IR asymptotics remain unchanged and the UV asymptotics (i.e. $\theta\to\pm\infty$ for $\varepsilon_\pm(\theta)$) simply becomes
\begin{equation}\label{UVasympboost}
    \varepsilon_\pm(\theta)\to\left(\beta\mp\nu-\frac{\pi \sigma}{6(\beta\pm\nu)}\tilde{c}_{\mp}\right)E_\pm(\theta)+w_\pm(\theta).
\end{equation}
Observe that the NLIEs are actually the same as \eqref{nlieGGE1} upon replacing $\beta\mp\nu$ with $\tilde{\beta}_\pm=\beta\mp\nu-\frac{\pi \sigma}{6(\beta\pm\nu)}\tilde{c}_{\mp}$. In a boosted state, which has $w_\pm(\theta)=0$,  this allows us to compute the scaling function $\tilde{c}_\pm(\beta,\nu)$ explicitly, which now depends on $\beta$ and $\nu$ due to the deformation. To be more precise, we have
\begin{equation}
    \tilde{c}_\pm(\beta,\nu)=\frac{\beta\mp\nu}{\tilde{\beta}_\pm}\frac{3\tilde{\beta}_\pm M}{2\pi^2}\int_{-\infty}^\infty\dd \theta e^{\pm\theta} L_\pm(\theta)=\frac{\beta\mp\nu}{\tilde{\beta}_\pm}\frac{c}{2},
\end{equation}
which can be solved simply:  
\begin{equation}\label{extscaling}
    \tilde{c}_\pm(\beta,\nu)=\frac{3(\beta^2-\nu^2)}{\pi\sigma}\left(1-\sqrt{1-\frac{\pi\sigma c}{3(\beta^2-\nu^2)}}\right).
\end{equation}
For  the thermal case $\nu=0$, this gives the free energy $f=-\pi\tilde{c}(\beta)/(6\beta^2)$, which is in agreement with the one obtained from solving the Burgers equation \eqref{free-energy}. The explicit $\tilde{c}_\pm(\beta,\nu)$ allows us to write the NLIEs for the boosted state in a more transparent way
\begin{equation}
    \varepsilon_\pm(\theta)=\tilde{\beta}_{\pm}(\beta,\nu) E_\pm(\theta)-T\star L_\pm(\theta),
\end{equation}
where
\begin{equation}
    \tilde{\beta}_{\pm}(\beta,\nu)=\frac{\beta\mp\nu}{2}\left(1+\sqrt{1-\frac{\pi\sigma c}{3(\beta^2-\nu^2)}}\right).
\end{equation}
We next compute the dressed quantities. At finite temperatures (or in GGEs), thermodynamic quantities associated with quasi-particles get dressed due to their interactions with the background. In general, when $T$ is symmetric, the dressing operation to a function $f_\pm(\theta)$ is defined by $f^\dr_\pm(\theta)=f_\pm(\theta)+[T n_\pm f_\pm^\dr](\theta)+[\tilde{T}_{\pm\mp} n_\mp f_\mp^\dr](\theta)$, where we used the integral operator representation $[Tf](\theta)=\int\dd\theta'T(\theta,\theta')f(\theta')$. Those that are relevant for our purpose are $E_\pm^\dr,p^\dr_\pm$ and $(E'_\pm)^\dr,(p'_\pm)^\dr$, the former of which can be conveniently obtained by $E^\dr_\pm(\theta)=\p \varepsilon_\pm(\theta)/\p\beta$ and  $p^\dr_\pm(\theta)=-\p \varepsilon_\pm(\theta)/\p\nu$. We first notice that $E_\pm^\dr=(p'_\pm)^\dr$ and $p^\dr_\pm=(E'_\pm)^\dr$, since $E_\pm=p'_\pm$ and $p_\pm=E'_\pm$. From \eqref{nlieGGE2} we then have
\begin{align}
    (p'_\pm)^\dr(\theta)&=\frac{\p\tilde{\beta}_\pm}{\p\beta}E_\pm(\theta)+[T\star n(p'_\pm)^\dr](\theta)\n
   (E'_\pm)^\dr(\theta)&=-\frac{\p\tilde{\beta}_\pm}{\p\nu}E_\pm(\theta)+[T\star n(E'_\pm)^\dr](\theta),
\end{align}
which actually implies that
\begin{equation}\label{veffdef}
    v^\eff_\pm(\theta)=\frac{(E'_\pm)^\dr(\theta)}{ (p'_\pm)^\dr(\theta)}=-\frac{\p\tilde{\beta}_\pm}{\p\nu}\bigg/\frac{\p\tilde{\beta}_\pm}{\p\beta}.
\end{equation}
This is a crucial observation in applying GHD to \ttbar-deformed CFTs; contrary to the common belief, the deformation does not induce curvature to the dispersion relation, but merely modifies the light-cone velocity. In particular, in the partitioning protocol this immediately implies that the NESS profile always consists of two {\it contact discontinuities} which can be thought of as two entropically stable shocks. We will study the phenomenon in detail later. We also note that in a thermal state $v^\eff$ can be readily evaluated and reads
\begin{equation}\label{veff}
     v^\eff_\pm(\theta)=\pm\sqrt{1-\frac{\pi\sigma c}{3\beta^2}}.
 \end{equation}
Observe the superluminal behavior when $\sigma<0$ in the thermal case.
 
Interestingly, in fact $v^\eff$ can be written solely in terms of the energy densities $\rho_\pm$, which has an important consequence in hydrodynamics. To see this, we first note from \eqref{centralcharge} and \eqref{veffdef} that
  \begin{equation}
      v^\eff_\pm=\frac{\pm1+\sigma \rho_\mp v^\eff_\mp}{1+\sigma \rho_\mp},
  \end{equation}
  which can be solved as
  \begin{equation}
      v^\eff_\pm=\frac{\pm1+\sigma(\rho_+-\rho_-)}{1+\sigma(\rho_++\rho_-)},
  \end{equation}
  where $\rho_\pm$ are chiral energy densities $\rho_\pm=\int\frac{\dd\theta}{2\pi}p'_\pm(\theta)n_\pm(\theta)E^\dr_\pm(\theta)$. We stress that this is true in any situation.
\section{Hydrodynamics of $T\bar{T}$-deformed CFTs and the reversible cellular automaton 54}
\begin{figure}[h!]
\centering
\includegraphics[width=8cm]{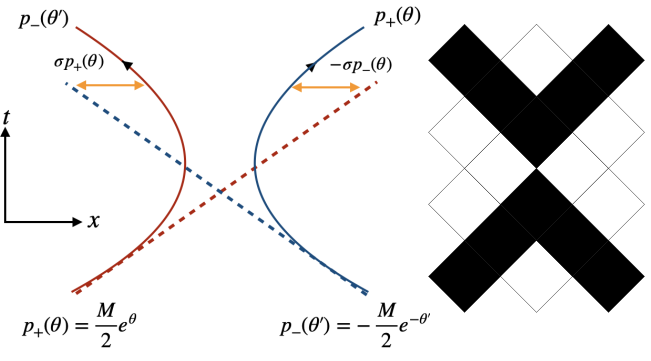}
\caption{Scattering between a right and left mover in a $T\bar{T}$-deformed CFT, and in the RCA 54.}
\label{ttbarscat}
\end{figure}

The fact that $v^\eff_\pm(\theta)$ does not depend on $\theta$ is of paramount importance in applying GHD to \ttbar-deformed CFTs. Let us recall that the GHD equation in terms of the particle density $\rho^\mathrm{p}_\pm(\theta)=(p'_\pm)^\dr(\theta)n_\pm(\theta)/(2\pi)$, where $n_\pm(\theta)=1/(1+e^{\varepsilon_\pm(\theta)})$ is the occupation function, reads
\begin{equation}
    \p_t\rho^\mathrm{p}_\pm(\theta)+\p_x(v^\eff_\pm(\theta)\rho^\mathrm{p}_\pm(\theta))=0.
\end{equation}
Since $v^\eff_\pm(\theta)$ is independent of $\theta$, the equation for each $\theta$ is actually decoupled, wherefore we can solve the equation for each $\theta$ independently. More importantly, this implies that the hydrodynamic equation for the chiral energy densities $\rho_\pm$ is {\it closed}, i.e. the equation evolves self-consistently. The hydrodynamic equation then reads
\begin{equation}\label{RCA54}
\p_t\rho_\pm+\p_x(v^\eff_\pm\rho_\pm)=0,\quad v^\eff_\pm=\frac{\pm1+\sigma(\rho_+-\rho_-)}{1+\sigma(\rho_++\rho_-)}.
\end{equation}
Observe that the equation is invariant under the scaling $\sigma\mapsto a\sigma$ and $\rho_\pm\mapsto\rho_\pm/a$. Using this scaling we can always define new densities $\rho_{\mathrm{sol},\pm}:=\sigma\rho_\pm$, with which the equation becomes the GHD equation for a celebrated integrable cellular automaton model called Rule 54 chain (RCA54) \cite{Friedman2019}. 
RCA 54 (also known as the Floquet Fredrickson-Andersen model) is one of the simplest  interacting systems \cite{Bobenko1993}. It comprises left and right moving solitons propagating at a constant velocity $\pm 1$. Upon scattering the two solitons undergo a constant phase shift. Despite this apparent simplicity the model provided a playground for understanding transport properties \cite{Gopalakrishnan2018,Friedman2019,klobas2019time}, and operator spreading \cite{PhysRevB.98.060302,PhysRevLett.122.250603} in integrable classical and quantum systems.
As mentioned earlier, remarkably, the GHD equation for \ttbar-deformed CFTs and the RCA 54 are exactly the same with the identification $\rho_{\mathrm{sol},\pm}=\sigma\rho_\pm$. Namely, the energy densities in \ttbar-deformed CFTs, $\rho_\pm$, can be interpreted as the densities of right/left moving solitons in the RCA 54 with an appropriate scaling by $\sigma$. Furthermore, on the Euler scale,  the particle current in the RCA 54 can be identified as the energy current $\langle j_E\rangle=\rho_+-\rho_-$ in \ttbar-deformed CFTs. A crucial distinction between \ttbar-deformed CFTs and the RCA 54 is that in the latter the particles cannot be arbitrarily close due to the exclusion principle, inducing the hard core interaction. Since \ttbar-deformed CFTs are field theories, there is no such restrictions, and the hard-core self-scattering terms are replaced by the {\it soft} self-scattering terms (i.e. scatterings that depend on $\theta)$. While this has an impact on thermodynamics, it does not affect the hydrodynamics, since the same particle species in both \ttbar-deformed CFTs and the RCA 54 never scatter due to the same velocities\footnote{Therefore the notion of scattering matrices between the same species should be taken with a grain of salt; they are obtained by the massless limit of some massive theories, and ultimately justified by the fact that they reproduce the expected properties of CFTs \cite{Fendley1994}}. See Fig.~\ref{ttbarscat} for a comparison between a scattering in a \ttbar-deformed CFT and the RCA 54.

GHD equations are usually labeled by the continuous parameter $\theta$. However, we have just observed above that when specializing to energy densities, a coupled two-component hydrodynamics suffices to predict the energy dynamics in \ttbar-deformed CFTs. Such reduction to a finite-component system of GHD in a particular situation has been previously known to occur only when integrable models whose ground states are Luttinger liquids are (locally) in zero-entropy states consisting of multi-Fermi seas \cite{PhysRevLett.119.195301}. It is important to bear it in mind that the mechanism of these two reductions are completely different; the one here is {\it exact} in the sense that these two equations are enough to describe the dynamics of a fluid, which never suffers from shocks (gradient catastrophe), as solutions of these equations are given by contact discontinuities, which can be thought of as stable shocks that are not accompanied by any entropy production. Another difference is that the equation is for the energy density, while in the zero-entropy case the equation can be written down for the dynamical Fermi points, which therefore could be used to compute the dynamics of other observables.
\section{NESS and transport coefficients in $T\bar{T}$-deformed CFTs}
To study the partitioning protocol in $T\bar{T}$-deformed CFTs, we make use of the results from GHD, and in particular the equation in terms of $\rho_\pm$. A Riemann problem of a hyperbolic system of this type was in fact worked out before \cite{PhysRevLett.95.204101}; the solution simply consists of two contact discontinuities along $\xi=-v^\eff_L$ and $\xi=v^\eff_R$, where $\xi=x/t$, emanating in space-time (see Appendix for the detail). A contact discontinuity can be considered as a stable shock in the sense that there is no entropy production across the jump, and  is realized as a consequence of a merger of a shock and a rarefaction wave. In fact, the existence of a contact discontinuity is intimately related to the fact that some eigenvalue of the lineralization matrix $A_i^{\,\,j}=\p \langle j_j\rangle/\p\langle q_i\rangle$ is {\it linearly degenerate}, i.e. there exists an eigenvalue $v^\eff_i$ such that $\sum_k (R^{-1})_k^{\,\,i}\p v^\eff_i/\p\langle q_k\rangle=0$, where $RAR^{-1}=\mathrm{diag}(v^\eff)$. Indeed, it has been known that fluids of integrable systems constitute a large class of totally linear degenerate systems, which are characterized by the fact that all the eigenvalues of the matrix $A$ are linearly degenerate \cite{El2010}. Here by directly solving a Riemann problem in the reduced GHD equation for $\rho_\pm$, we can confirm this peculiar property of integrable systems. Consequently we obtain a rather simple profile of a Riemann problem in a \ttbar-deformed CFT; a current-carrying steady state emerges in between $-v^\eff_R<\xi<v^\eff_L$, outside of which the state corresponds to the asymptotic baths. In the two-temperatures partitioning protocol starting from two baths with temperatures $T_R$ and $T_L$, which are subject to $T_L,T_R<T_\mathrm{H}$, we can in fact derive the closed forms of the NESS currents for any \ttbar-deformed CFTs. Solving \eqref{RCA54} with the initial condition $\rho_\pm(x,0)=\rho_{\pm,L}\vartheta(-x)+\rho_{\pm,R}\vartheta(x)$ (see Appendix:A for the derivation), where $\rho_{\pm,R/L}$ are energy densities evaluated with respect to temperatures $T_{R/L}$, the energy and momentum NESS currents $\langle j_E\rangle=\rho_+-\rho_-$ and $\langle j_P\rangle=\rho_+v^\eff_+-\rho_-v^\eff_-$ are given by
\begin{align} \label{cicurrent2}
    \langle j_E\rangle_\mathrm{NESS}&=\frac{\pi c }{12}e_{RL}\left(\frac{1}{\tilde{\beta}^2_L}-\frac{1}{\tilde{\beta}^2_R}\right),\n
    \langle j_P\rangle_\mathrm{NESS}&=\frac{\pi c }{12}e_{RL}\left(\frac{1}{\tilde{\beta}^2_L}+\frac{1}{\tilde{\beta}^2_R}-\frac{\pi c\sigma}{6\tilde{\beta}^2_L\tilde{\beta}^2_R}\right),
\end{align}
where
\begin{equation}
    e_{RL}=\frac{1}{1-\left(\frac{\pi\sigma c}{12}\right)^2\frac{1}{\tilde{\beta}_R^2\tilde{\beta}_L^2}}
\end{equation}
and $\tilde{\beta}_{R,L}=\tilde{\beta}_\pm(\beta_{R,L},0)$. Some comments are due.  First, note that when taking the undeformed limit $\sigma\to0$, \eqref{cicurrent2} reduces to the aforementioned CFT result, as expected. Second, from this expression, it is clear that the contributions to the NESS current stem from the right (resp. left) movers that are thermalized with respect to the left (resp. right) bath, which is exactly the same as in pure CFTs.  A critical difference from the CFT formula is that \eqref{cicurrent2} cannot be written as $f(\beta_L)- f(\beta_R)$, which is a consequence of the scatterings between the left and the right movers.  Notice, however, that such mixing is absent in the first order in $\sigma$ in accordance with the perturbative result \cite{Bernard_2016}. In general, we expect that the contributions from the deformation can be divided into two parts: thermodynamic contributions that are linear in $\sigma$ and scattering (dynamical) contributions that depend on higher powers of $\sigma$ (a similar observation was also made in \cite{Cardy_2016}). The fact that the effect of scatterings is encoded through the coefficient $e_{RL}$ also reminds us of how NESS currents are modified in pure CFTs in the presence of an impurity at $x=0$ \cite{Bernard_2015}.
 Second, in the two-temperatures partitioning protocol, the local equilibrium state always takes a form of a boosted thermal state. In particular, the NESS is specified by the effective temperature $T=\sqrt{T_RT_L}$ with the boost parameter  $\tanh\nu=(\beta_L-\beta_R)/(\beta_L+\beta_R)$, which again coincides with the structure of the NESS in the undeformed CFTs. An even more important impact of the dynamical contributions can be observed in the momentum spreading, which admits diffusive corrections that are absent in the pure CFTs or at the first order in $\sigma$  \cite{Bernard_2016}. Later we shall compute the exact momentum diffusion constant for \ttbar CI. 

To summarize, the only effect induced by the \ttbar-deformation in the partitioning protocol in the \ttbar-deformed CFTs is the modification of the light cone velocities, whose magnitude could be larger/smaller than $1$ depending on the sign of the deformation parameter $\sigma$. It is interesting to compare our finding with the previous perturbative analysis performed in \cite{Bernard_2016}, in which it was concluded that, at least at the leading order in $\sigma$, there were regions in space-time where the exact profile of a fluid could not be determined by hydrodynamic consideration. This is because the derivative corrections can be completely absorbed by redefining fluid variables, and thus entropic condition (i.e. a condition that ensures the stability of shocks) cannot be imposed to single out a weak solution. We speculate that, in \ttbar-deformed CFTs, such undetermined regions will shrink as one includes higher order terms in $\sigma$, and eventually vanish, resulting in the simple profile obtained below.
 \begin{figure}[h!]
\centering
\includegraphics[width=7cm]{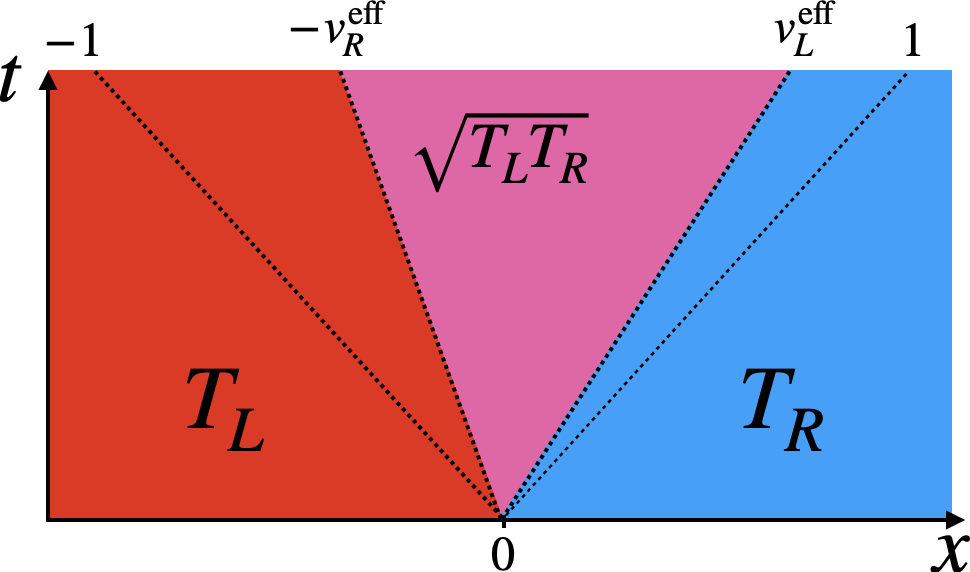}
\caption{The partitioning protocol starting with two heat baths with temperatures $T_L$ and $T_R$ in a $T\bar{T}$-deformed CFT. The whole space-time is divided by the modified light cones with velocities $-v^\eff_R$ and $v^\eff_L$. In the stationary region a current-carrying NESS current emerges with the effective temperature $T=\sqrt{T_LT_R}$.}
\label{ttbar_riemann}
\end{figure}
It is clear by construction that the modification to the pure CFT result is caused by the $R-L$ scattering introduce by the \ttbar-deformation. From the holographic point of view, however the effect of the deformation incarnates in a completely different way. Namely there is a dual picture of the \ttbar-deformation: one is the introduction of the CDD factor while the another one is the field-dependent coordinate change, the latter of which is exploited in holography. To be more precise, particles in a \ttbar-deformed theory behave like free particles in the new coordinate, which incidentally hints at the picture proposed in the context of GHD \cite{cardy2020toverline}. Later, we shall employ similar point of view in order to compute the NESS currents using the AdS/CFT correspondence.
Next we turn to Drude weights. In order to compute the Drude weights exactly, we can use the general expression derived in the scope of the GHD \cite{SciPostPhys.3.6.039}
\begin{equation}\label{drude}
    D_{ij}=\sum_{a=\pm}\int_\mathbb{R}\dd \theta\rho_a(\theta)(1-n_a(\theta))(v^\eff_a(\theta))^2h^\dr_{a,i}(\theta)h^\dr_{a,j}(\theta),
\end{equation}
where $\rho_a(\theta)=n_a(\theta)\rho^\tot(\theta)$ with $\rho^\tot(\theta)=(p'_a)^\dr(\theta)/(2\pi)$. In general it is hard to explicitly carry out the integration, but in thermal equilibrium, we can compute the energy Drude weight from the free energy in a boosted state by appealing to the fact that the energy current is conserved in \ttbar-deformed CFTs: $D_{EE}=-\left.\p^2 f/\p \nu^2\right|_{\nu\to0}$. Furthermore, using the fact that $p^\dr_\pm(\theta)=v^\eff_\pm E^\dr_\pm(\theta)$ in a boosted state, we can actually relate the energy and the momentum Drude weights as $D_{PP}=D_{EE}(v^\eff)^2$. The results turn out to be surprisingly simple, and read
\begin{equation}
    D_{EE}=\frac{\pi c}{3\beta^3}\frac{1}{v^\eff},\quad D_{PP}=\frac{\pi c}{3\beta^3}v^\eff,
\end{equation}
where
\begin{equation}\label{thermalveff}
    v^\eff=\sqrt{1-\frac{\pi\sigma c}{3\beta^2}}.
\end{equation}
Note that both quantities admit a perturbative expansion in $\sigma$ and are also UV finite so long as $T<T_\mathrm{H}$. 
Through the holographic computation, we shall later provide another derivation of the NESS currents as well as Drude weights which agree perfectly, whereby validating these universal formulae as well as serving as a new nontrivial confirmation of the deformed AdS/CFT correspondence in (1+1)-dimension.

Finally, we turn to the momentum diffusion, which in fact can also be explicitly computed for the \ttbar-deformed CFTs. To do so, we employ the formula for the diffusion constants in integrable systems obtained within the framework of GHD
\begin{align}\label{physicalons}
    \mathfrak{L}_{ij}&=\frac{1}{2}\sum_{a,b}\int\dd\theta\dd\lambda\chi_a(\theta)\chi_b(\lambda)|v^\eff_a(\theta)-v^\eff_b(\lambda)|\n
    &\quad\times\left(\frac{T^\dr_{ba}(\lambda,\theta)h^\dr_{i,b}(\lambda)}{\rho^\tot_{b}(\lambda)}-\frac{T^\dr_{a\beta}(\theta,\lambda)h^\dr_{i,a}(\theta)}{\rho^\tot_{a}(\theta)}\right)\n
    &\quad\times\left(\frac{T^\dr_{ba}(\lambda,\theta)h^\dr_{j,b}(\lambda)}{\rho^\tot_{b}(\lambda)}-\frac{T^\dr_{ab}(\theta,\lambda)h^\dr_{j,a}(\theta)}{\rho^\tot_{a}(\theta)}\right),
\end{align}
where $\chi_a(\theta)=\rho_a(\theta)(1-n_a(\theta))$ is the quasi-particle susceptibility. We are in particular interested in the momentum diffusion, which turns out to be given by a simple expression in \ttbar-deformed CFTs\begin{widetext}
\begin{equation}
    \mathfrak{L}_{PP}=8\pi^2(v^\eff)^3\int\dd\theta\dd\lambda\chi(\theta)\chi(\lambda)(T^\dr_{+-}(\theta,\lambda)+T^\dr_{-+}(\lambda,\theta))^2.
\end{equation}
\end{widetext}
A crucial observation that facilitates the computation is that in fact even after the dressing operation the phase shift $T_{\pm\mp}$ is factorized: $T^\dr_{\pm\mp}(\theta,\lambda)=-\sigma p^\dr_\pm(\theta)p^\dr_\mp(\lambda)/(2\pi)$. To see this, note first that $T^\dr_{\pm\mp}(\theta,\lambda)$ satisfies
\begin{align}\label{Tdr}
    T^\dr_{\pm\mp}(\theta,\lambda)&=-\frac{\sigma}{2\pi}\left(p_\pm(\theta)+[T^\dr np_\pm](\theta)\right) p_\mp(\lambda) \n
    &\quad+[T^\dr_{\pm\mp} nT](\theta,\lambda),
\end{align}
where we used the relation $TnT^\dr=T^\dr nT$ (integral operator representation is implied).
Likewise the dressing equation for $p^\dr_\pm$ can be written as
\begin{equation}
    p^\dr_\pm(\theta)=p_\pm(\theta)+[T^\dr_{\pm\mp}np_\mp](\theta)+[T^\dr np_\pm](\theta),
\end{equation}
which allows us to rewrite \eqref{Tdr} in an alternative way
\begin{align}\label{Tdr2}
    T^\dr_{\pm\mp}(\theta,\lambda)&=-\frac{\sigma}{2\pi}\left(p^\dr_\pm(\theta)+[T^\dr_{\pm\mp} np_\mp](\theta)\right) p_\mp(\lambda) \n
    &\quad+[T^\dr_{\pm\mp} nT](\theta,\lambda).
\end{align}
Recalling that
\begin{equation}
     p^\dr_\pm(\theta)=\left(1-\frac{\sigma}{2\pi}[p^\dr_\pm np_\mp](\theta)\right)p_\pm(\theta)+[p^\dr_\pm nT](\theta),
\end{equation}
It is then clear that $T^\dr_{\pm\mp}(\theta,\lambda)=-\sigma p^\dr_\pm(\theta)p^\dr_\mp(\lambda)/(2\pi)$ satisfies \eqref{Tdr2}, which we assume as the unique solution (a Fredholm integral equation usually admits a unique solution). Using this expression for $T^\dr_{\pm\mp}$, and the fact that at the thermal equilibrium $E^\dr_+=E^\dr_-$ and $p^\dr_+=-p^\dr_-$, we end up with
\begin{equation}
    \mathfrak{L}_{PP}=\frac{\sigma^2}{2}\left(\frac{\pi c}{3\beta^3}\right)^2\frac{1}{v^\eff}.
\end{equation}
Notice that it goes to zero in the undeformed limit $\sigma\to0$, and in particular, the effect of deformation starts at the second order in $\sigma$, which is in agreement with the previous discussion. We can therefore interpret the diffusion as a fingerprint of scatterings between the left and the right movers.

In the next two sections, we explain how the general structure in \ttbar-deformed CFTs shows up in two illustrative examples: the critical Ising model and the Liouville CFT. In particular we shall provide the exact momentum diffusion constant in \ttbar CI.
\section{$T\bar{T}$-deformed critical Ising model ($T\bar{T}$CI)}
 Without  $T\bar{T}$-deformation, the critical Ising model is a free CFT with the central charge $c=1/2$ and trivial S-matrices, which makes the model much more amenable to analytical treatments than interacting CFTs. The deformation induces a nontrivial interaction between two chiral modes controlled by the two-body S-matrix $S(p,q)=e^{\ii\sigma pq}$, where $p=p_+(\theta)$ is the energy of the right mover, while $q=-p_-(\theta)$ is the energy of the left mover \cite{Caselle2013}. In \ttbar CI we shall use a parameterization in terms of $p$ and $q$, which turns out to be useful. TBA of the \ttbar CI is governed by the pseudo-energy $\varepsilon_\pm$ \cite{Caselle2013}
\begin{equation}\label{pseudo}
    \varepsilon_\pm(p)=\beta p-\int_0^\infty\dd q T^\mathrm{T}(p,q)L_\mp(q),
\end{equation}
with $L_\pm(p)=\log(1+e^{-\varepsilon_\pm(p)})$. 
In thermal equilibrium \eqref{pseudo} can be solved explicitly, yielding $\varepsilon_\pm(p)=\varepsilon p$ with
\begin{equation}\label{psuedoext}
     \varepsilon=\frac{1}{2}\left(\beta+\sqrt{\beta^2-\frac{\pi \sigma}{6}}\right).
\end{equation}

 Another ingredients needed for our purpose are the dressed functions of quasi-momentum $p$. If $h_\pm(p)$ is invariant under the reparameterization $p\mapsto f(p)$ (e.g. $h_\pm(p)=E_\pm(p)$), i.e. $h_\pm(p)$ is a scalar, then its dressed version can be obtained from $h_\pm^\dr=h_\pm+T^\mathrm{T}n_\mp h^\dr_\mp$. If instead $h_\pm(p)\dd p$ is invariant under such repramerization (e.g. $h_\pm(p)=E'_\pm(p)$), i.e. $h_\pm(p)$ is a vector, then $h_\pm^\dr$ can be obtained as $h_\pm^\dr=\pm h_\pm+Tn_\mp h^\dr_\mp$.
 Notice that we distinguish between $T$ and $T^\mathrm{T}$, which is important in the parameterization in terms of $p$ and $q$. Remarkably, in $T\bar{T}$CI such dressing transformations can be carried out exactly, for quantities such as $E_\pm(p)=p$ and $p_\pm(p)$, yielding rather simple expressions: $E^\dr_\pm(p)=e_-p$, $p^\dr_\pm(p)=\pm e_+p$, $(E'_\pm)^\dr(p)=\pm e_+$, $(p'_\pm)^\dr(p)=e_-$, where $e_+=\varepsilon/\beta$ and $e_-=\varepsilon/(2\beta-\varepsilon)$. In particular the effective velocity $v^\eff$ reads
 \begin{equation}\label{veffising}
     v^\eff_\pm(p)=\pm\sqrt{1-\frac{\pi\sigma}{6\beta^2}},
 \end{equation}
 which is in accordance with \eqref{veff}. 

\section{\ttbar-deformed Liouville CFT}
While \ttbar CI exhibits representative features of \ttbar-deformed CFTs, some of its properties are subject to the fact that the underlying CFT is free. To incorporate the effect of interaction, we shall deal with the \ttbar-deformed Liouville CFT (\ttbar LC), and in particular obtain the energy Drude weight of \ttbar LC, confirming \eqref{drudeCFT}. The reason to study \ttbar LC is threefold: first, the Liouville CFT is one of the most well-studied interacting CFTs that is {\it not} a minimal model. Its rich structure also allows us to obtain a plethora of analytic results, e.g. the DOZZ formula for the three-point function \cite{DORN199239,ZAMOLODCHIKOV1996577}. Second, the central charge of the Liouville CFT is always greater than 25, and can actually be arbitrarily large by tuning the model parameter. This is particularly useful in view of the comparison with the holographic CFTs. Third, most importantly for our purpose, the TBA-like formulation of the Liouville CFT as per Bazhanov-Lukyanov-Zamolodchikov is actually known, and was reported in the unpublished paper by Alexei Zamolodchikov, entitled ``Generalized Mathieu equation and liouville TBA'' \cite{zamolodchikov2012quantum}. Let us recall some of the essential and somewhat anomalous properties of the Liouville TBA.

The non linear integral equation (NLIE) that describes the Liouville CFT in a boosted Gibbs state $\rho\sim e^{-\beta H+\nu P}$ is given by
\begin{equation}\label{ltba}
    \varepsilon_\pm(\theta)=(\beta\mp\nu) E_\pm(\theta)-T\star L_\pm(\theta),
\end{equation}
The phase shift $T(\theta)$ is in fact the same as the one for the massive sinh-Gordon model \cite{Zamolodchikov_2006}
\begin{equation}
    T(\theta)=\frac{1}{2\pi}\frac{4\cosh\theta\cos\frac{\pi a}{2}}{\cosh2\theta+\cos(\pi a)},
\end{equation}
with $a=(1-b^2)/(1+b^2)$. Here, $b$ is the model parameter that can be restricted to $0<b\leq 1$. What distinguishes the Liouville TBA from the sinh-Gordon TBA is the asymptotics
\begin{equation}\label{asymp}
    \varepsilon_\pm(\theta)\to\pm4PQ\left(\theta+\log\frac{(\beta\mp\nu) M}{2\pi}\right)-2C(P)
\end{equation}
when $\theta\to\mp\infty$, and where $P$ is an arbitrary positive parameter and $Q=b+b^{-1}$. $C(P)$ is fixed by the NLIE \eqref{ltba} itself. The asymptotics of the opposite limit $\theta\to\pm\infty$ is $\varepsilon_\pm(\theta)\to\beta E_\pm(\theta)$ as in the sinh-Gordon model. The central charge of the system is given by the scaling function \eqref{centralcharge}. A variant of dilogarithmic computation shows that we can explicitly carry out the integration, obtaining $\tilde{c}_\pm(\beta,\nu)=(1+24P^2)/2$, hence $\tilde{c}(\beta,\nu)=1+24P^2$ (for the detail, see Appendix). The actual central charge for the Liouville CFT is then obtained by setting $P=Q/2$, which gives $c_L=1+6Q^2$. Notice that the value is always greater than $c=25$, which occurs at the self-dual point $b=1$.

Now, let us turn on the \ttbar-deformation. We can proceed exactly as in the general case. For instance, we can obtain  the free energy  as well as the scaling function that exactly match with \eqref{free-energy} and \eqref{extscaling} with $c=c_L$.We emphasize that at the level of NESS currents the \ttbar LC and the \ttbar-deformed holographic CFTs behave in same way for any value of $c$, without having to take the large $c$ limit.
\section{Holography and transport}

The holographic correspondence relates a CFT in $d$ dimensions to a gravity theory in $d+1$ dimensions. Correlators of CFT operators can be obtained by studying the fluctuations of fields in an asymptotically AdS spacetime (referred to as "the bulk", as opposed to the boundary of AdS where the CFT lives). 
In particular, the stress-energy tensor is dual to the graviton flucutations, which satisfy the equations of motion derived from the Einstein-Hilbert action 
\begin{equation}
    S = \frac{1}{16 \pi G_N} \int_{\mathbb{R}^{d+1}} d^{d+1}x \sqrt{g} (R - 2 \Lambda) \,,
\end{equation}
where $\Lambda = -\frac{d(d-1)}{2 \ell^2}$ is the cosmological constant, and $\ell$ is the radius of curvature of the background AdS solution around which the fields fluctuate. A generic solution admits the Fefferman-Graham expansion $ds^2 = \ell^2 \frac{d\rho^2}{4 \rho^2}+\frac{1}{\rho}\gamma_{\mu\nu} dx^\mu dx^\nu$ with $\gamma_{\mu\nu} = \gamma^{(0)}_{\mu\nu}(x)+\rho \gamma^{(2)}_{\mu\nu}(x)+ {\cal O}(\rho^2)$. In the holographic dictionary this geometry correspond to a state in the CFT that lives on the boundary at $\rho=0$ with a background metric $\gamma^{(0)}$, and $\gamma^{(2)}$ encoding the expectation value of $T_{\mu\nu}$ in this state. For applications to transport properties we are usually interested in the thermal state, which is dual to a black brane geometry. 
An important point is that because gravity is described universally by the Einstein-Hilbert action, at the lowest order in derivatives, the properties of stress-energy tensor correlators, and quantities derived from them, can in some cases be found independently of other details of the CFT and are therefore universal (for instance, the shear viscosity of a conformal holographic plasma is one such quantity). In general, however, the dual gravity theory may contain higher-derivative corrections (e.g. higher powers of the Riemann tensor) which are not universal. Neglecting these higher-order terms corresponds to a strong-coupling limit, and considering the classical theory in the bulk, by neglecting quantum corrections, corresponds to the limit of large central charge of the CFT. There can also be other fields coupled to gravity. In this paper we will consider only the simplest model described by pure gravity.

The holographic dual of the partitioning protocol was considered in \cite{Bhaseen:2015aa}. It was shown there that the solution for the NESS generated by two baths at different temperature is a boosted black brane, with boost and temperature determined by the initial conditions $T_L, T_R$. They found, in the case of $d=2$, the result for the NESS energy current in agreement with \cite{bernard2016conformal}, as well as the extension to higher dimensional CFTs. As mentioned in the introduction, in the $2d$ case with an unbroken conformal symmetry we don't expect any particular insight from holography as far as transport properties are concerned.  
\section{$T\bar{T}$-deformed holographic CFTs}\label{sec:holographic}

We discuss here the implementation of the partitioning protocol in the deformed holographic correspondence. 
We follow closely \cite{Guica:2019nzm}, where the gravity solutions relevant for our problem are given. It is well-known that the most general solution of pure gravity in 3d with a cosmological constant, and a flat boundary metric at infinity, is given by the Ba\~nados geometry \cite{Banados:1998gg}
\begin{align}\label{Banados}
    ds^2 &= \ell^2 \frac{d\rho^2}{4 \rho^2} + \frac{du dv}{\rho} +{\cal L} (u) \, du^2 +\nonumber\\ &+ \bar {\cal L}(v) \, dv^2 + \rho {\cal L}(u) \bar {\cal L}(v) \, du dv \,.
\end{align}
Here $\rho$ is the radial direction, as in section III, $u,v$ are light-cone coordinates at the boundary, $\ell$ is the $AdS$ radius, related to the central charge by $c = \frac{3 \ell}{2 G}$. The solution depends on two arbitrary functions $\cL(u), \bcL(v)$ that describe the left and right moving excitations of the CFT, and are in fact related to the expectation value of the stress-energy tensor: $\cL = 8 \pi G \ell \langle T_{uu} \rangle, \bcL = 8 \pi G \ell \langle T_{vv}\rangle $. When 
$\cL = \bcL=0$, the bulk geometry is pure AdS and is dual to the vacuum of the CFT; the metric seen by the CFT is $du \, dv$. A particularly important class of solutions is the rotating BTZ black hole, dual to the boosted thermal ensemble  $e^{-\beta(H - \Omega P)}$; its mass and angular momentum are related to $\cL, \bcL$, which are constants in this case\footnote{In the gravity literature the space coordinate is usually taken to be an angle, therefore the notion of the angular momentum is used. We are considering the case where the space is a line and consequently the angular momentum becomes the linear momentum.}. 

Following the prescription of \cite{McGough:2016lol,Guica:2019nzm}, the $T \bar{T}$-deformed solution must have a flat induced metric on the cutoff surface $\rho= \mu$. One might expect that as a consequence the metric at infinity will not be flat, which would take us outside of the Ba\~nados metrics. However, it turns out that 
$\sqrt{\gamma} R(\gamma)$ (where $\gamma$ is the induced metric on a slice of constant $\rho$) is preserved under the flow of the deformation, implying that also the metric at infinity is flat. Importantly, in 2d any two flat metrics can be related by a coordinate transformation. The solution is obtained by the following change of coordinates
\begin{equation}\label{coordchange}
U = u + \mu \int^v \bar{\cal L}(v') dv' \,, \quad V = v + \mu \int^u {\cal L}(u') du' \,. 
\end{equation}
One can check that in the new coordinates, the metric \eqref{Banados} at the cutoff surface is simply $dU dV$. 
The simplicity of the solution is related to the special properties of the $T \bar{T}$-deformation: its equivalence to a state-dependent change of coordinates has been emphasized in \cite{Dubovsky:2017cnj,Cardy:2019qao}.
The explicit form of the solution cannot be obtained in the general case, but for our purpose we only need to consider (piecewise) constant $\cL,\bcL$ and in this case we can easily invert the relations \eqref{coordchange}. Note however that even though \eqref{coordchange} is linear in $\mu$, the metric in the new variables is given by a non-linear, and non-polynomial expression. 

Once we have the solution for the boosted thermal ensemble, we are in a position to implement the partitioning protocol, as was done in the case of undeformed CFT in \cite{Bhaseen:2015aa}. At time $t=0$ we join two different solutions on the left and right; for this we have to consider piece-wise constant functions 
\begin{equation} 
{\cal L}(u) = {\cal L}_L \theta(-u) + {\cal L}_R \theta(u) \,, \quad \bar {\cal L}(v) =\bar  {\cal L}_L \theta(-v) + \bar {\cal L}_R \theta(v) \,.
\end{equation}
The solution develops two shock waves that propagate along the lines $u=0$ and $v=0$. Inverting the coordinate change in this case, the shock waves trajectories become, at $t>0$, 
\begin{equation}
x= - \frac{1+ \mu \bar {\cal L}_L}{1-\mu \bar {\cal L}_L} t \,, \quad x= \frac{1+ \mu {\cal L}_R}{1-\mu {\cal L}_R} t \,,
\end{equation}
where $x,t$ are related to the light-cone coordinates as $U = x+t, V= x-t$. 
These last equations generalize the perturbative result for the speed of the shock in \cite{Bernard_2016} to all orders in $\sigma$. Expressing the speed in terms of temperature, $v=\sqrt{1+\frac{4 \pi^2 \ell^2 \mu}{\beta^2}}$ and this agrees  with \eqref{effective-v} once we use  the relation between the cutoff and the deformation parameter \eqref{mu-sigma} that we repeat here for convenience: 
\begin{equation}
    \mu = - \frac{c}{12 \pi \ell^2} \sigma \,.
\end{equation}
As explained in the appendix, the prescription formulated in terms of mixed boundary conditions makes sense for either sign of $\mu$. The case of positive $\mu$, which corresponds to a finite cutoff, lead to superluminal propagation, as observed in \cite{Marolf:2012dr}. Notice also that the classical gravity computation requires us to scale $\sigma \sim 1/c$, if we want to keep the cutoff fixed in terms of the AdS radius. 

The non-equilibrium steady state that develops between the two shock waves has a form of the general solution, with $\cL = \cL_R, \bcL = \bcL_L$. One can then simply read off the NESS energy and momentum currents from the 
expressions for the stress-energy tensor \eqref{EMtensor}. If we consider the zero-momentum initial states, namely $\cL_L = \bcL_L,\cL_R=\bcL_R$, we find
\begin{align}  \label{JE-hol}
    \langle j_E \rangle_\mathrm{NESS} &=  \frac{1}{8 \pi G \ell} \frac{\cL_L - \cL_R}{1- \mu^2 \cL_L \cL_R} \,, \\
    \langle j_P \rangle_\mathrm{NESS} &= \frac{1}{8 \pi G \ell} \frac{\cL_L + \cL_R + 2 \mu \cL_L \cL_R}{1- \mu^2 \cL_L \cL_R} \,.
\end{align}
We recognize the same structure as in \eqref{cicurrent2}, and we can check that the formulae agree precisely once we express the solution's parameters in terms of temperature (see \eqref{parameters}):
\begin{equation}
    \sqrt{\cL} = \frac{\beta  \left(\sqrt{\frac{4 \pi ^2 l^2 \mu}{\beta ^2}+1}-1\right)}{2 \pi  l \mu } = \frac{\pi \ell}{\tilde{\beta}_\pm(\beta,0)} \,.
\end{equation}
As explained in the introductory section on energy transport, we can extract the Drude weights from the NESS currents. 
Both, the energy and the momentum Drude weights are then given in terms of thermodynamic quantities: 
\begin{equation}
 D_{EE} =  \frac{\mathtt{e}+\mathtt{p}}{\beta} \,, \quad D_{PP} = \left( \frac{\mathtt{p}}{\mathtt{e}} \right)^2 D_{EE},
\end{equation}
which hold at all orders in the deformation parameter. 

In holography, the presence of a horizon in the bulk is the mechanism by which a theory can exhibit dissipation at finite temperature: matter simply disappears behind the horizon; the rate of absorption can be related to the diffusion constant. 
In 2d CFT the conformal symmetry prevents dissipation from happening even at finite temperature; as is well-known, the finite temperature theory is related to the one at zero temperature by a conformal mapping. The holographic manifestation of this fact is that pure gravity in 3d has no propagating degrees of freedom: any solution is locally pure AdS. The $T \bar{T}$-deformation breaks conformal symmetry, and one could expect that diffusive behavior emerge in gravity already at the classical level. This, however, is not the case: the only effect of the $T \bar{T}$-deformation at the classical level is the modification of the effective speed of propagation. In fact, the deformed solution appears to be, at the classical level, equivalent to a CFT in a deformed geometry,  so that there is effectively no mixing of left/right movers. The diffusion constant should presumably arise from the quantum gravity corrections, which goes beyond the scope of the present work. 


\section{Diffusion from conformal perturbation theory}

In this section we show that the diffusion constants, or the Onsager matrix, can be obtained very generally, to the second order in the deformation, using CFT methods. 
The Onsager matrix is given, in linear response theory, by the formula \eqref{Onsager}. As we already mentioned, only the diagonal momentum component of the matrix can be non-vanishing because of relativistic invariance. This is the term coming from the correlator $ \langle j_P j_P \rangle  = \langle T_{xx} T_{xx} \rangle$. It is easy to see that a correlator $ \langle T_{tt} \,j_i \rangle$, for any current $j_i$, cannot give a finite contribution in \eqref{Onsager} because $T_{tt}$ is a conserved charge density. Therefore we can substitute $ \langle j_P j_P \rangle$ with $ \langle (T_{xx}-T_{tt})(T_{xx}-T_{tt})
\rangle$. This is the correlator of the trace of $T$, and an important property of the $T \bar{T}$ deformation is that the 
trace satisfy the relation 
\begin{equation}\label{tracerel}
    \mathrm{tr}T = - \sigma \, \mathrm{det}(T_{\mu\nu}) \,,
\end{equation}
as can be checked, for instance, in the holographic formulation, \eqref{trace}. At leading order we find $ \mathrm{tr}T(z,\bar z) = \frac{\sigma}{\pi^2} T(z,\bar z) \bar T(z,\bar z)$. The two point function of the trace is then a four point function evaluated in the underformed CFT, and it factorizes between the left and right movers; in real time we have
\begin{widetext}
\begin{equation}
    \langle \mathrm{tr}\,T(x,t) \mathrm{tr}\,T(0,0) \rangle = \frac{\sigma^2}{\pi^4} \frac{c^2}{4} \frac{1}{\sinh^4(\frac{\pi}{\beta}(x+t))\sinh^4(\frac{\pi}{\beta}(x-t))} \,. 
\end{equation}
\end{widetext}
The right-hand side has the form of the finite-temperature correlator of a scalar operator of dimension 4. The retarded correlator of an operator of conformal dimension $(h, \bar h)$ is \cite{Oshikawa_2002}
\begin{align}
    G^R(\omega,q) = & - \sin(2 \pi h) (2\pi T)^{2(h+ \bar h -1)}  \times \\
    & B(h-i \frac{\omega+q}{4 \pi T},1-2h) B(\bar h-i \frac{\omega-q}{4 \pi T},1-2 \bar h) \nonumber
\end{align}
where $B(x,y)=\frac{\Gamma(x) \Gamma(y)}{\Gamma(x+y)}$ is the Euler beta function. The correlator is singular at integer values of the dimension, so we expand it around $h=\bar h=2$ and keep the finite part. Finally, recalling the fluctuation-dispersion relation between the retarded $G^R$ and symmetrized correlator $C$,  $\mathrm{Im} G^R(\omega,q) = (1-e^{-\beta \omega}) C(\omega,q)$, we can evaluate the Onsager coefficient: 
\begin{equation}
     \mathfrak{L}_{PP} = {\sigma^2 \over 2} \frac{\pi^2 c^2}{9 \beta^6} \,.
\end{equation}
This agrees with the result \eqref{LPP} to order $\sigma^2$.  
Notice that the diffusion constant, given by the Onsager coefficient divided by the susceptibility, is of order $\sigma^2 c$, therefore it is suppressed as $1/c$ in the classical limit $c \to \infty$ with $\sigma c$ fixed. To see it in holography would require a computation of quantum gravity effects.

\section{Conclusion}
In the present paper we presented the results on energy and momentum transport in the $T\bar{T}$-deformed CFTs by employing integrability and holography. First, we solved the partitioning protocols, with the two sides of the system prepared at different temperatures,  exactly, using both integrability and holography. Although these methods have presumably disconnected ranges of validity, the results turn out to agree, yielding universal closed form expressions for the energy and momentum NESS currents in \ttbar-deformed CFTs with the arbitrary central charge. This, in turn, also allowed us to compute the energy and momentum Drude weights. Furthermore, we calculated the exact momentum Onsager matrix, which is related to the momentum diffusion constant, in generic \ttbar-deformed CFTs using the idea of GHD. The result is also supported by conformal perturbative expansion up to the second order in the deformation parameter. These computations of the momentum Onsager matrix are consistent with the fact that it vanishes in holographic CFTs in the leading order of $\frac{1}{c}$.

Our results have far reaching consequences from a number of different perspectives. First of all, we calculate the corrections of the famous Stefan Boltzmann law by considering integrable deformations of the conformal field theory. Secondly, we made an identification between the $T\bar{T}$ dynamics and RCA 54. And lastly, the matching of the Drude weights between GHD and holography represent a first verification of the $T\bar{T}$-deformed gauge/gravity correspondence on the level of dynamics.

There are many questions that the present manuscript opens up from the point of view of holography. In general $T\bar{T}$-deformed CFTs offer a playground where the dynamical aspects of holography can be tested by exact methods. This includes entanglement dynamics, operator spreading (which is related to the Lyapunov exponent), and additional features of transport. In particular, we have only focused on Euler scale hydrodynamics, which, however, admits corrections. First order corrections are directly connected to the diffusion constant and since we found no diffusion at the classical level, it should arise as a consequence of quantum corrections to the gravity. Such corrections may be computable using the effective theory formulation of \cite{Cotler:2018zff}.  Notice also that the scaling $\sigma \sim 1/c$, required for the gravity dual, is the same scaling used in \cite{Aharony:2018vux} to derive evolution equations for the correlation functions; 
generalizing this method to finite temperature  would provide an alternative way to derive $1/c$ corrections.

 Another aspect that deserves further consideration is that of the observed universality. Namely, we found universal expressions for the NESS currents and Drude weights. It would be interesting to understand this universality from the viewpoint of the holographic dual, by considering different models in which gravity can interact with other fields.
It would be very interesting also to consider more general current-current deformations like $J{\bar T}$ \cite{Guica:2017lia}, which break Lorentz invariance and would lead to energy diffusion.
 In general, we would like to establish a full connection at a level of generalized Gibbs ensembles, which would require the inclusion of higher-order conserved charges,
that are dual to higher-spin fields in the bulk. This would allow us to derive the full expressions for the Drude weights and diffusion constant in terms of the normal modes of GHD \cite{Medenjak:2019fke,doyon2019diffusion} from holography. 


Another exciting future direction is offered by the connection between $T\bar{T}$-deformed CFTs and RCA 54 on hydrodynamic level. In fact, this observation is consistent with the finding in a recent work \cite{cardy2020toverline} in which Cardy and Doyon observe that generically a \ttbar-deformation can be thought of as a deformation that introduces a ``width'' to constituent particles. To make the connection valid even at the level of thermodynamics, we however need to introduce hard-core $R-R$ and $L-L$ scatterings in \ttbar-deformed CFTs. A natural question to ask is if we could concoct another (irrelevant) operator that amounts to such scatterings. A somewhat related direction was pushed in \cite{PhysRevLett.124.200601} where the authors studied the consequences of adding the most generic CDD factors to scattering matrices in integrable quantum field theories. A careful look, however, shows that such deformations cannot yield scatterings between the same species. In general, relating the two theories from microscopics represents an exciting future perspective, which might lead to the holographic CA.

\section{Acknowledgements}
We are indebted to useful comments and discussions with Olalla Castro-Alvaredo, Benjamin Doyon and Monica Guica. TY is also extremely grateful to Stefano Negro for many useful exchanges, and in particular for suggesting us to study the Liouville CFT. TY thanks Benjamin Doyon for sending his article \cite{cardy2020toverline} to us before publication. GP acknowledges the funding from the NYU-PSL research project "Holography and Quantum Gravity" (reference ANR-10-IDEX-0001-02 PSL).  

\vspace{0.2cm}
\bibliographystyle{apsrev4-1}
\bibliography{TC}

\appendix

\pagebreak


\section{Riemann problem in a \ttbar-deformed CFT}
In this section we solve the two-temperature partioning protocol in a \ttbar-deformed CFT, which can be formulated as a Riemann problem in RCA54 where the solition densities are now replaced by the energy densities $\rho_\pm$ in a \ttbar-deformed CFT. To reiterate, the GHD equation reads
\begin{equation}\label{dghd}
    \begin{pmatrix}
    \p_t\rho_+ \\
    \p_t\rho_-
    \end{pmatrix}
    +
    \p_x\begin{pmatrix}
  v^\mathrm{eff}_+&0 \\
   0&v^\mathrm{eff}_-
    \end{pmatrix} \begin{pmatrix}
    \rho_+ \\
    \rho_-
    \end{pmatrix}=0,
    \end{equation}
with 
\begin{equation}\label{riemannveff}
    \begin{pmatrix}
    v^\mathrm{eff}_+ \\
    v^\mathrm{eff}_-
    \end{pmatrix}
    =\frac{1}{1+\sigma(\rho_++\rho_-)}
    \begin{pmatrix}
  1+\sigma(\rho_+-\rho_-) \\
   -1+\sigma(\rho_+-\rho_-)
    \end{pmatrix}.
\end{equation}
It is in fact easier to work with another fluid coordinate $u=\rho_++\rho_-, v=\rho_+-\rho_-$ by realizing the following relation
\begin{align}
    \rho_+v^\mathrm{eff}_++\rho_-v^\mathrm{eff}_-&=\rho_+-\rho_- \\
    \rho_+v^\mathrm{eff}_+-\rho_-v^\mathrm{eff}_-&=\rho_++\rho_--\frac{4\sigma\rho_+\rho_-}{1+\sigma(\rho_++\rho_-)}.
\end{align}
In terms of $u$ and $v$, \eqref{dghd} can be recast into
\begin{align}
    \p_tu+\p_xv&=0 \\
    \p_tv+\p_x\Big(u-\frac{\sigma(u^2-v^2)}{1+\sigma u}\Big)&=0,
\end{align}
which can also be represented as
\begin{equation}\label{dghd2}
    \begin{pmatrix}
    \p_t u \\
    \p_tv
    \end{pmatrix}
    +
    J \begin{pmatrix}
    \p_xu \\
    \p_xv
    \end{pmatrix}=0,\quad J=\begin{pmatrix}
 0&1 \\
   \frac{1-\sigma^2v^2}{(1+\sigma u)^2}&\frac{2\sigma v}{1+\sigma u}
    \end{pmatrix}.
    \end{equation}
    Note that the eigenvalues of the Jacobian is $v^\mathrm{eff}_\pm$ as it should be.
    
    We want to solve a general Riemann problem with an initial condition
    \begin{equation}
        u(x,0)=\begin{dcases*}
        u_L & $x<0$\\
        u_R & $x>0$
        \end{dcases*},\quad  v(x,0)=\begin{dcases*}
        v_L & $x<0$\\
        v_R & $x>0$
        \end{dcases*}.
    \end{equation}
    Let us start with looking for a two-shocks solution that is natural to exist on the physical ground. To find it, we first solve the Rankine-Hugoniot condition for the left moving shock connecting the left state $u_L,v_L$ and the right state that is to be determined $u_*,v_*$. Denoting the shock speed $\xi_1$, the equations we have to solve are
    \begin{align}
        v_L-v_*&=\xi_1(u_L-u_*) \label{rhl1}\\
        u_L-\frac{\sigma(u_L^2-v_L^2)}{1+\sigma u_L}-u_*+\frac{\sigma(u_*^2-v_*^2)}{1+\sigma u_*}&=\xi_1(v_L-v_*).\label{rhl2}
    \end{align}
    After some laborious calculations, quite surprisingly, it turns out that the shock speed $\xi_1$ does not depend on $u_*$ nor $v_*$, and simple reads
    \begin{equation}
        \xi_1=\frac{\sigma v_L\pm1}{\sigma u_L+1}=v^\mathrm{eff}_\pm(u_L,v_L).
    \end{equation}
    The sign can be determined as follows: when both densities are zero $\rho_{\pm,L}=0$ (i.e. $v_L=u_L=0$), we want the shock to propagate to the left, thus we choose the minus sign. 
    
   In the exactly same manner, we can perform a similar analysis for the second shock connecting the left state $u_*,v_*$  and the right state $u_R,v_R$ with the shock speed $\xi_2$. The set of equations to be satisfied are
   \begin{align}
        v_R-v_*&=\xi_2(u_R-u_*) \label{rhr1}\\
        u_R-\frac{\sigma(u_R^2-v_R^2)}{1+\sigma u_R}-u_*+\frac{\sigma(u_*^2-v_*^2)}{1+\sigma u_*}&=\xi_2(v_R-v_*),\label{rhr2}
    \end{align}
   yielding
    \begin{equation}
        \xi_2=\frac{\sigma v_R\pm1}{\sigma u_R+1}=v^\mathrm{eff}_\pm(u_R,v_R).
    \end{equation}
    We will choose the plus sign here as well for the similar reason above. Plugging $\xi_1,\xi_2$ into \eqref{rhl1} and \eqref{rhr1}, we have
    \begin{equation}
        \begin{pmatrix}\label{ness}
     u_* \\
    v_*
    \end{pmatrix}=\frac{1}{\sigma(\xi_2-\xi_1)}\begin{pmatrix}
     \xi_1-\xi_2+2 \\
    \xi_2+\xi_1
    \end{pmatrix}
    \end{equation}
    which gives
    \begin{equation}
        v^\mathrm{eff}_-(u_*,v_*)=\xi_1,\quad v^\mathrm{eff}_+(u_*,v_*)=\xi_2.
    \end{equation}
    This implies a remarkable feature of the fluid. Namely, the Lax condition for the stability of shocks are satisfied with equality:
    \begin{align}
         v^\mathrm{eff}_-(u_L,v_L)&\geq\xi_1\geq v^\mathrm{eff}_-(u_*,v_*), \n v^\mathrm{eff}_+(u_*,v_*)&\geq\xi_2\geq v^\mathrm{eff}_+(u_R,v_R).
    \end{align}
This class of wave is called {\it contact discontinuity}, and it turns out that both the left and the right moving shocks are contact discontinuities. 

We are then finally in the position to obtain the explicit steady state profile $(u_*,v_*)$, which, according to \eqref{ness}, is
 \begin{equation}
        \begin{pmatrix}\label{ness2}
     u_* \\
    v_*
    \end{pmatrix}=\frac{1}{\sigma(v^\eff_{+,R}-v^\eff_{-,L})}\begin{pmatrix}
    v^\eff_{-,L}-v^\eff_{+,R}+2 \\
   v^\eff_{+,R}+v^\eff_{-,L}
       \end{pmatrix}.
    \end{equation}
Since $v$ is nothing but the energy current in the language of the \ttbar-deformed CFTs, we immediately obtain \eqref{cicurrent2} by plugging \eqref{riemannveff} into \eqref{ness2}.

\section{Central charge in the Liouville CFT}
Here we explicitly compute the central charge of the Liouville CFT at the self-dual point; the computation away from the point is completely analogous. To facilitate the calculation, we set $\beta M=2\pi$ without loss of generality. Let us start with introducing two functions $L_0(\theta)$ and $L_1(\theta)$ defined by
\begin{align}
    L_0(\theta)&=2PQ\log(1+e^{-2\theta}) \\
    L_1(\theta)&=\frac{C}{2}(1-\tanh\theta).
\end{align}
Their convolutions with $T(\theta)$ can be easily computed, reading
\begin{align}
    f_0(\theta)&:=T*L_0(\theta)=2PQ\log\left(1+e^{-2\theta}+2e^{-\theta}\cos\frac{\pi a}{2}\right) \\
    f_1(\theta)&:=T*L_1(\theta)=\frac{C}{2}\left(1-\frac{\sinh\theta}{\cosh\theta+\cos\frac{\pi a}{2}}\right).
\end{align}
With these functions, we can rewrite the TBA equation \eqref{ltba} as
\begin{equation}
    \varepsilon(\theta)=\pi e^\theta-f_0(\theta)-f_1(\theta)-[T*(L_R-L_0-L_1)](\theta),
\end{equation}
which is also useful for numerics in that its asymptotics is properly controlled. The object of interest here is the central charge
\begin{equation}\label{ccharge}
   c_P= \frac{6}{\pi}\int\dd\theta e^\theta\log(1+e^{-\varepsilon(\theta)}),
\end{equation}
where we suppressed the subscript $+$ of $\varepsilon(\theta)$. A crucial difference between the standard dilogarithm computations in TBA and the one we spell out below is that some of the integrations that appear in the computation might not be convergent, and can be made finite only when the divergence is canceled out by a divergence in another integral. First notice from differentiating \eqref{ltba} with respect to $\theta$ that
\begin{equation}\label{epderiv}
    \pi e^\theta=(1-Tn)\p_\theta\epsilon.
\end{equation}
Replacing $e^\theta$ in \eqref{ccharge} with this, we proceed the calculations as follows
\begin{align}
    c&=\frac{6}{\pi^2}\int_{-\infty}^\infty\dd\theta L(1-Tn)\p_\theta\varepsilon \n
    &=\frac{6}{\pi^2}\int_{\varepsilon(-\infty)}^\infty\dd\varepsilon\log(1+e^{-\varepsilon})\n
    &\quad+\frac{6}{\pi^2}\int_{-\infty}^\infty\dd\theta L(\theta)\int_{-\infty}^\infty\dd\theta'T(\theta-\theta')L'(\theta').
\end{align}
Let us focus on the double integral in the second term. Importantly these two integrals do not commute due to the divergence from the $\theta$-integration, hence we cannot appeal to the standard trick to change the order of integrations and replace the $\theta$-integration using the TBA equation \eqref{ltba}. To compute the integration with better control, we shall isolate the integration that causes divergence by rewriting it as
\begin{align}\label{bldb1}
    &\int_{-\infty}^\infty\dd\theta L(\theta)\int_{-\infty}^\infty\dd\theta'T(\theta-\theta')L'(\theta')\n
    &=\int_{-\infty}^\infty\dd\theta (L(\theta)-\tilde{L}(\theta))\int_{-\infty}^\infty\dd\theta'T(\theta-\theta')(L'(\theta')-\tilde{L}(\theta')) \n
    &\quad+\int_{-\infty}^\infty\dd\theta L(\theta)\int_{-\infty}^\infty\dd\theta'T(\theta-\theta')\tilde{L}'(\theta')\n
    &\quad+\int_{-\infty}^\infty\dd\theta \tilde{L}(\theta)\int_{-\infty}^\infty\dd\theta'T(\theta-\theta') L'(\theta')\n
    &\quad-\int_{-\infty}^\infty\dd\theta \tilde{L}(\theta)\int_{-\infty}^\infty\dd\theta'T(\theta-\theta')\tilde{L}'(\theta'),
\end{align}
where we introduced $\tilde{L}(\theta)=L_0(\theta)+L_1(\theta)$.
The two integrations in the first line commute safely now, thus making use of \eqref{ltba} one can massage them to get
\begin{widetext}
\begin{align}\label{bldb2}
    &\int_{-\infty}^\infty\dd\theta (L(\theta)-\tilde{L}(\theta))\int_{-\infty}^\infty\dd\theta'T(\theta-\theta')(L'(\theta')-\tilde{L}'(\theta')) \n
    &=\int_{-\infty}^\infty\dd\theta' (L'(\theta')-\tilde{L}'(\theta'))\left[\pi e^{\theta'}-\epsilon(\theta')-\int_{-\infty}^\infty\dd\theta T(\theta-\theta')\tilde{L}(\theta)\right]\n
    &=-\pi\int_{-\infty}^\infty\dd\theta e^\theta L(\theta)-\int_{-\infty}^\infty L'(\theta)\epsilon(\theta)-\int_{-\infty}^\infty\dd\theta' \tilde{L}'(\theta')\int_{-\infty}^\infty\dd\theta T(\theta-\theta')L(\theta)\n
    &\quad+\int_{-\infty}^\infty\dd\theta'L'(\theta')\int_{-\infty}^\infty\dd\theta T(\theta-\theta')  \tilde{L}(\theta)+\int_{-\infty}^\infty\dd\theta' \tilde{L}'(\theta')\int_{-\infty}^\infty\dd\theta T(\theta-\theta')\tilde{L}(\theta).
\end{align}
The second and the third terms in RHS of \eqref{bldb1} can be written just in terms of $\tilde{L}$ when combined with some of the terms in \eqref{bldb2}. To be precise, we have
\begin{align}
    &\int_{-\infty}^\infty\dd\theta L(\theta)\int_{-\infty}^\infty\dd\theta'T(\theta-\theta')\tilde{L}'(\theta')-\int_{-\infty}^\infty\dd\theta' \tilde{L}'(\theta')\int_{-\infty}^\infty\dd\theta T(\theta-\theta')L(\theta)\n
    &=\int_{-\infty}^\infty\dd\theta' \tilde{L}(\theta')\int_{-\infty}^\infty\dd\theta T(\theta-\theta')\tilde{L}'(\theta)-\int_{-\infty}^\infty\dd\theta' \tilde{L}'(\theta')\int_{-\infty}^\infty\dd\theta T(\theta-\theta')\tilde{L}(\theta)
\end{align}
and
\begin{align}
    &\int_{-\infty}^\infty\dd\theta \tilde{L}(\theta)\int_{-\infty}^\infty\dd\theta'T(\theta-\theta')L'(\theta')-\int_{-\infty}^\infty\dd\theta' L'(\theta')\int_{-\infty}^\infty\dd\theta T(\theta-\theta')\tilde{L}(\theta)\n
    &=\int_{-\infty}^\infty\dd\theta' \tilde{L}(\theta')\int_{-\infty}^\infty\dd\theta T(\theta-\theta')\tilde{L}'(\theta)-\int_{-\infty}^\infty\dd\theta' \tilde{L}'(\theta')\int_{-\infty}^\infty\dd\theta T(\theta-\theta')\tilde{L}(\theta).
\end{align}
Therefore merging everything together, we end up with
\begin{align}
    \int_{-\infty}^\infty\dd\theta L(\theta)\int_{-\infty}^\infty\dd\theta'T(\theta-\theta')L'(\theta')&=-\pi\int_{-\infty}^\infty\dd\theta e^\theta L(\theta)+\int_{\epsilon(-\infty)}^\infty\dd\epsilon\frac{\epsilon}{1+e^\epsilon}\n
    &\quad+\int_{-\infty}^\infty\dd\theta' \tilde{L}(\theta')\int_{-\infty}^\infty\dd\theta T(\theta-\theta')\tilde{L}'(\theta)\n 
    &\quad  -\int_{-\infty}^\infty\dd\theta' \tilde{L}'(\theta')\int_{-\infty}^\infty\dd\theta T(\theta-\theta')\tilde{L}(\theta),
\end{align}
implying that the central charge is now given by
\begin{align}
    \label{ccharge2}
    c&=1+\frac{3}{\pi^2}\left(\int_{-\infty}^\infty\dd\theta' \tilde{L}(\theta')\int_{-\infty}^\infty\dd\theta T(\theta-\theta')\tilde{L}'(\theta)-\int_{-\infty}^\infty\dd\theta' \tilde{L}'(\theta')\int_{-\infty}^\infty\dd\theta T(\theta-\theta')\tilde{L}(\theta)\right) \n
    &=1+\frac{3}{\pi^2}\int_{-\infty}^\infty\dd\theta\left[(L_0(\theta)+L_1(\theta))(f'_0(\theta)+f'_1(\theta))-(L'_0(\theta)+L'_1(\theta))(f_0(\theta)+f_1(\theta))\right],
\end{align}
\end{widetext}
where we noted that
\begin{equation}
   \frac{3}{\pi^2}\int_{\epsilon(-\infty)}^\infty\dd\epsilon\left[\log(1+e^{-\epsilon})+\frac{\epsilon}{1+e^\epsilon}\right]=1.
\end{equation}
So the task boils down to compute the single integration in \eqref{ccharge2}, which we shall denote by $I$. So far the computation is valid for any value of $Q$, but to illustrate how the remaining computation can be achieved, we again specialize to the self-dual case. It turns out that it is convenient to divide the integration into the term proportional to $P^2$, $PC$, and $C^2$. We first deal with the term that comes with $P^2$, which reads
\begin{widetext}
\begin{align}
    & 32P^2\int_{-\infty}^\infty\dd\theta\left[-\log(1+e^{-2\theta})\frac{1}{1+e^\theta}+2\log(1+e^{-\theta})\frac{1}{1+e^{2\theta}}\right] \n
    &=32P^2\int_{-\infty}^\infty\dd\theta\left[-\log(1+e^{2\theta})\frac{e^\theta}{1+e^\theta}+2\log(1+e^{\theta})\frac{e^{2\theta}}{1+e^{2\theta}}\right] \n
    &=32P^2\int_{-\infty}^\infty\dd\theta\left[-\left(2\theta+\log(1+e^{-2\theta})\right)\left(1-\frac{1}{1+e^\theta}\right)+2\left(\theta+\log(1+e^{-\theta})\right)\left(1-\frac{1}{1+e^{2\theta}}\right)\right] \n
    &=32P^2\int_{-\infty}^\infty\dd\theta\left[2\log(1+e^{-\theta})-\log(1+e^{2\theta})+2\theta\left(\frac{1}{1+e^\theta}-\frac{1}{1+e^{2\theta}}\right)\right] \n
    &\quad-32P^2\int_{-\infty}^\infty\dd\theta\left[-\log(1+e^{-2\theta})\frac{1}{1+e^\theta}+2\log(1+e^{-\theta})\frac{1}{1+e^{2\theta}}\right].
\end{align}
Hence the term is simplified and can be exactly computed to be
\begin{align}
    &32P^2\int_{-\infty}^\infty\dd\theta\left[-\log(1+e^{-2\theta})\frac{1}{1+e^\theta}+2\log(1+e^{-\theta})\frac{1}{1+e^{2\theta}}\right] \n
    &=32P^2\int_{0}^\infty\dd\theta\left[2\log(1+e^{-\theta})-\log(1+e^{2\theta})+2\theta\left(\frac{1}{1+e^\theta}-\frac{1}{1+e^{2\theta}}\right)\right] \n
    &=128P^2\int_0^\infty\dd\theta\,\theta\left(\frac{1}{1+e^\theta}-\frac{1}{1+e^{2\theta}}\right) \n
    &=96P^2\int_0^\infty\dd\theta\,\theta\frac{1}{1+e^\theta} \n
    &=8\pi^2P^2,
\end{align}
where in the second line we noted that the integrand is an even function of $\theta$.

Next we move on to the term with $PC$, which is given by
\begin{align}
   &2PC\int_{-\infty}^\infty\dd\theta\bigg[-\log(1+e^{-2\theta})\frac{\dd}{\dd\theta}\tanh\frac{\theta}{2}+2\log(1+e^{-\theta})\frac{\dd}{\dd\theta}\tanh\theta\bigg] \n
    &\quad+4PC\int_{-\infty}^\infty\dd\theta\bigg[-2\frac{1}{1+e^\theta}(1-\tanh\theta)+2\frac{1}{1+e^{2\theta}}(1-\tanh\frac{\theta}{2})
    \bigg].
\end{align}
It turns out that each integration is identically zero. For instance, the first line becomes
\begin{align}
   & \int_{-\infty}^\infty\dd\theta\bigg[-\log(1+e^{-2\theta})\frac{\dd}{\dd\theta}\tanh\frac{\theta}{2}+2\log(1+e^{-\theta})\frac{\dd}{\dd\theta}\tanh\theta\bigg] \n
   &=\int_{-\infty}^\infty\dd\theta\bigg[-\tanh\frac{\theta}{2}+2\tanh\theta\frac{1}{1+e^\theta}\bigg]\n
   &=\int_{-\infty}^\infty\dd\theta\frac{\tanh\frac{\theta}{2}}{\cosh\theta}\n
   &=0
\end{align}
\end{widetext}
Likewise, one can also show that the second line also vanishes.

Finally we are left with the term proportional to $C^2$, but it is immediately to realise that it is zero:
\begin{align}
    &C^2\int_{-\infty}^\infty\dd\theta\left[-(1-\tanh\theta)\frac{1}{2\cosh^2\frac{\theta}{2}}+\left(
    1-\tanh\frac{\theta}{2}\right)\frac{1}{\cosh^2\theta}\right]\n
    &=C^2\int_{-\infty}^\infty\dd\theta\left(\frac{1}{\cosh^2\theta}-\frac{1}{2\cosh^2\frac{\theta}{2}}\right)\n
    &=0.
\end{align}
Therefore we conclude that the central charge $c$ in the self-dual case is nothing but
\begin{equation}
    c=1+24P^2
\end{equation}
with $P=Q/2=1$.
\section{Black hole solutions in the deformed correspondence}

We give some details on the gravity solutions used for the partitioning protocol. As explained in the main text, the 
solution for a black hole with finite temperature and momentum is obtained by \eqref{Banados}, with constant $\cL,\bcL$, with the change of coordinates \eqref{coordchange}. 
The explicit result is, in terms of the Fefferman-Graham expansion given in section III, 
\begin{align}\label{deformed-BH}
    \gamma^{(0)}_{\mu\nu} dx^\mu dx^\nu&  = \frac{(dU - \mu \bcL dV)(dV- \mu \cL dU)}{(1-\mu^2 \cL \bcL)^2} \nonumber \\
     \gamma^{(2)}_{\mu\nu} dx^\mu dx^\nu & = \frac{(1+\mu^2 \cL \bcL)}{(1-\mu^2 \cL \bcL)^2} (\cL dU^2+\bcL dV^2)\nonumber  \\
     & - \frac{4 \mu \cL \bcL dU dV}{(1-\mu^2 \cL \bcL)^2}  \\
      \gamma^{(4)}_{\mu\nu}  & = \cL \bcL  \gamma^{(0)}_{\mu\nu} \nonumber
\end{align}
The dual energy-momentum tensor is given by the Brown-York tensor on the $\rho=\mu$ slice, with the addition of the usual holographic counterterm \cite{deHaro:2000vlm}
\begin{align}
    T_{ij} & = T_{ij}^{BY} - \frac{1}{8 \pi G \ell} \gamma_{ij}(\rho=\mu)  \\
    & = \frac{1}{8 \pi G \ell} \frac{\cL dU^2+\bcL dV^2 + 2 \mu \cL \bcL dU dV}{1-\mu^2 \cL \bcL} \,.
\end{align}

Fixing the metric at $\rho=\mu$ means fixing $g= \gamma^{(0)}+\mu \gamma^{(2)}+\mu^2 \gamma^{(4)}$. Recalling that $\gamma^{(2)} \sim \langle T \rangle$, 
and the fact that $\gamma^{(4)} \sim ((\gamma^{(0)})^{-1}\gamma^{(2)})^2$, this can be reinterpreted as a mixed non-linear boundary condition between the metric and the energy-momentum tensor. With this interpretation one is not limited to positive values of $\mu$. 

The solution has an horizon at $\rho= 1/\sqrt{\cL \bcL}$. One can compute the Hawking temperature and the (angular) velocity by expanding near the horizon 
in suitable coordinates such that the horizon is at $r=0$ and the metric has the form $ds^2 = dr^2 - \frac{4\pi^2}{\beta^2} r^2 dt^2+ A(r) (dx - \Omega(r) dt)^2$. One finds
\begin{align}\label{parameters}
    \beta = &\frac{\pi \ell}{2}\left( \frac{1}{\sqrt{\cL}}+ \frac{1}{\sqrt{\bcL}} \right) (1- \mu \sqrt{\cL \bcL}) \,,\\
    \Omega = & \frac{\sqrt{\cL}-\sqrt{\bcL}}{\sqrt{\cL}+\sqrt{\bcL}} \,.
    \end{align}
We see that $\cL = \bcL$ corresponds to the non-boosted, or non-rotating, case. It is useful to give the expressions for the components of the 
energy-momentum tensor in $x,t$ coordinates: 
\begin{align}\label{EMtensor}
    T_{tt} &= \frac{1}{8 \pi G \ell} \frac{\cL + \bcL -2 \mu \cL \bcL}{1- \mu^2 \cL \bcL} \,, \nonumber \\
    T_{xt} & = \frac{1}{8 \pi G \ell} \frac{\cL - \bcL}{1- \mu^2 \cL \bcL} \,, \\
    T_{xx} & = \frac{1}{8 \pi G \ell} \frac{\cL + \bcL + 2 \mu \cL \bcL}{1- \mu^2 \cL \bcL} \,.  \nonumber
\end{align}    
We can find the equation of state for the pressure $\mathtt{p} = T_{xx}$ in terms of energy density $\mathtt{e} = T_{tt}$ and momentum density $\mathtt{j} = T_{xt}$: 
\begin{equation}
    \mathtt{p} = \frac{\mathtt{e} - 8 \pi G \ell \mu \mathtt{j}^2}{1-8 \pi G \ell \mu \mathtt{e}} \,.
\end{equation}
This is equivalent to the trace relation 
\begin{equation} \label{trace}
    \mathrm{tr} \, T = \mathtt{p} - \mathtt{e} = 8 \pi G \ell \mu \, \mathrm{det} T_{\mu\nu} \,.
\end{equation}
At zero momentum density, $\mathtt{j}=0$, we can express the pressure as 
\begin{equation}
    \mathtt{p} = \frac{1}{8 \pi G \ell} \frac{\cL}{1-\mu \cL} = \frac{\sqrt{1+\frac{4 \pi^2 \ell^2}{\beta^2}\mu} -1 }{8 \pi G \ell \mu} \,.
\end{equation}
The pressure is equal to the free energy, so comparing with \eqref{free-energy} we can find the relation between the 
cutoff radius and the deformation parameter: 
\begin{equation}\label{mu-sigma}
\mu = - \frac{\sigma c}{12 \pi \ell^2} \,.
\end{equation}

In order to compute the NESS current we have just to identify the relevant parameters for the solution. As stated in the main text,
the steady state that forms between the two shock waves is described by \eqref{deformed-BH} with parameters $\cL = \cL_R, \bcL = \bcL_L $. 
When we compute the NESS energy current, we can take the initial states to have zero boost, so $\cL_L = \bcL_L, \cL_R=\bcL_R$. Then the energy current density $j_E=T_{xt}$ is given by \eqref{EMtensor} with these parameters, which gives \eqref{JE-hol}; integrating between the position of the two shock waves we have the total current ${\cal J}_E(t) = \int dx \, j_E(x,t)$:
\begin{equation}
    \langle {\cal J}_E \rangle = \frac{1}{8 \pi G \ell} \frac{ {\cal L}_L-  {\cal L}_R}{1-\mu^2  {\cal L}_L   {\cal L}_R} \left( \frac{1+ \mu {\cal L}_L}{1-\mu {\cal L}_L}+\frac{1+ \mu {\cal L}_R}{1-\mu {\cal L}_R} \right)
\end{equation}
 
The Drude weight is obtained as 
\begin{equation}\label{energyDrude}\begin{aligned}
D_{EE} & = \frac{1}{2} \left(\frac{\partial}{\partial \beta_R} - \frac{\partial}{\partial \beta_R} \right)  {\cal J}_E |_{\beta_R=\beta_L}  \\
& = \frac{1}{2} \frac{\partial {\cal L}}{\partial \beta} \left(\frac{\partial}{\partial {\cal L}_R} - \frac{\partial}{\partial {\cal L}_L}\right) {\cal J}_E |_{{\cal L}_R={\cal L}_L} \\
& = \frac{\pi \ell}{2 G \beta^3}  \frac{1- \mu {\cal L}}{1+ \mu {\cal L}}  = \frac{\mathtt{e}+\mathtt{p}}{\beta} \,.\\
\end{aligned}\end{equation}
For the momentum Drude weight, we need to have a bias of angular velocity, but we could set $\beta_L=\beta_R$ in the initial state. However this does not give a simple relation 
in terms of $\cL,\bcL$, so we take arbitrary temperatures and set them equal only at the end. The momentum NESS current is the integral of $T_{xx}$ between the shock waves, and we need to subtract the expectation value in the initial state which is non-zero, unlike the case of the energy current. We find 
\begin{align}
    \langle {\cal J}_P \rangle & = \frac{1}{8\pi G \ell} \frac{{\cal L}_R+\bar {\cal L}_L + 2 \mu {\cal L}_R \bar {\cal L}_L}{1-\mu^2 {\cal L}_R \bar {\cal L}_L} \left(\frac{1+ \mu {\cal L}_R}{1-\mu {\cal L}_R}+\frac{1+ \mu \bar {\cal L}_L}{1-\mu \bar {\cal L}_L} \right) \nonumber \\
&    -\frac{1+ \mu {\cal L}_R}{1-\mu {\cal L}_R} j_{P,L} - \frac{1+ \mu \bar {\cal L}_L}{1-\mu \bar {\cal L}_L} j_{P,R}  \,.
\end{align}

The chemical potential coupled to the momentum density is $\nu = \beta \Omega$, so for the momentum Drude weight we have 
\begin{equation}
\begin{aligned}
D_{PP} = &  \frac{1}{ 2 \beta } (\frac{\partial {\cal J}_P}{\partial \Omega_L} |_{\beta_L}- \frac{\partial {\cal J}_P}{\partial \Omega_R}|_{\beta_R})|_{\beta_L=\beta_R,\Omega_L=\Omega_R=0} \\
= & \frac{\pi \ell}{2 G \beta^3} \frac{1+ \mu {\cal L}}{1- \mu {\cal L}} = \left( \frac{\mathtt{p}}{\mathtt{e}} \right)^2 D_{EE}\,.
\end{aligned}
\end{equation}

\end{document}